\documentclass[prl,twocolumn,tightenlines,nofootinbib,superscriptaddress]{revtex4-1}
\usepackage{bm,dcolumn,amsmath,graphicx,amsfonts,amssymb}

\usepackage{hyperref}
\usepackage{xcolor}
\definecolor{cite}{rgb}{0.,0.,0.85}   
\hypersetup{colorlinks,linkcolor={cite},citecolor={cite},urlcolor={cite}}

\newcommand{\abs}[1]{\ensuremath{\left |#1\right |}}
\newcommand{\braket}[1]{\ensuremath{\langle #1\rangle}}	

\renewcommand{\v}[1]{\ensuremath{\boldsymbol{#1}}}		

\def\d{\ensuremath{{\rm d}}}

\def\h{\ensuremath{\hbar}}

\newcommand{\s}{\ensuremath{\sigma}}
\renewcommand{\t}{\ensuremath{\tau}}
\renewcommand{\k}{\ensuremath{\kappa}}

\def\CT{\ensuremath{{\cal T}}}
\def\CR{\ensuremath{{\cal R}}}
\def\CL{\ensuremath{{\cal L}}}

\newcommand{\un}[1]{\ensuremath{\,{\rm{#1}}}} 

\newcommand{\sun}{\odot}

\usepackage[normalem]{ulem}
\definecolor{newc}{rgb}{0.,0.6,0.4}

\newcommand{\be}{\begin{equation}}
\newcommand{\ee}{\end{equation}}

\begin{document}

\title{Precision measurement noise asymmetry\\ and its annual modulation as a dark matter signature}

\author{Benjamin\ M.\ Roberts}\email[]{b.roberts@uq.edu.au}
\affiliation{School of Mathematics and Physics, The University of Queensland, Brisbane QLD 4072, Australia}
\affiliation{Department of Physics, University of Nevada, Reno, NV 89557, USA}
\author{Andrei Derevianko}\email[]{andrei@unr.edu}
\affiliation{Department of Physics, University of Nevada, Reno, NV 89557, USA}
\date{\today}

\begin{abstract}\noindent
Dark matter may be composed of self-interacting ultralight quantum fields that form macroscopic objects. 
An example of which includes Q-balls, compact non-topological solitons predicted by a range of theories that are viable dark matter candidates. 
As the Earth moves through the galaxy, interactions with such objects may leave transient perturbations in terrestrial experiments.
Here we propose a new dark matter signature: an asymmetry (and other non-Gaussianities) that may thereby be induced in the noise distributions of precision quantum sensors, such as atomic clocks, magnetometers, and interferometers.
Further, we demonstrate that there would be a sizeable annual modulation in these signatures due to the annual variation of the Earth velocity with respect to dark matter halo.
As an illustration of our formalism, we apply our method to 6 years of data from the atomic clocks on board GPS satellites and place constraints on  couplings for macroscopic dark matter objects with radii $R<10^4\un{km}$, the region that is otherwise inaccessible using relatively sparse global networks.
\end{abstract}

\maketitle

\section{Introduction}

Multiple astrophysical observations suggest that the ordinary (luminous or baryonic) matter  contributes only $\sim 5$\% to the total energy density budget of the Universe. Exacting the microscopic nature of the two other constituents, dark matter (DM) and dark energy remains a grand challenge to modern physics and cosmology.
DM is required for galaxy formations, while dark energy leads to the accelerated expansion of the Universe. The distinction between DM and dark energy can be formalized by treating them as cosmological fluids: they have different equations of state, DM is being pressureless,  while dark energy exerts negative pressure.  For further details the reader is referred to the cosmology textbooks, e.g., Ref.~\cite{weinberg-cosmology-2008} and reviews such as~\cite{Peebles2003,Bertone2005,Feng2010,Matarrese2011}. 

Exacting the microscopic nature of DM and its non-gravitational interaction with the standard model particles and fields is challenging. Indeed,  all the evidence for DM (galactic rotation curves, gravitational lensing, peaks in the cosmic microwave background spectra, etc)  comes from galactic scale (parsecs) observations.  The challenge lies  in extrapolating down from these scales to the laboratory scales and  a large number of theoretical  models can fit the observations.  All the theoretical constructs are guided by the cold dark matter paradigm that describes the large-scale structure formation  of the Universe~\cite{Blumenthal1984}. 

Despite composing the majority of matter in the universe, the microscopic nature of DM remains a mystery.
Most of the particle physics experiments so far have focused on weakly-interacting massive particles (WIMPs) with
$\sim$\,GeV\,--\,TeV masses.
Despite the extensive effort, there is no solid evidence for WIMPs in such ambitious large-scale experiments~\cite{XENON-1T-2017,Liu2017,Bertone2018}.
Besides WIMPs, there are a multitude of other DM candidates with masses that span many orders of magnitude.
Even if DM constituents are elementary particles, their masses can plausibly span 50 orders of magnitude: 
from $10^{-22} \un{eV}$ to $10^{28} \un{eV}$, with the lower bound coming from the requirements that their de Broglie wavelengths fit 
into dwarf galaxies, and the upper bound coming from the condition that they do not form black holes. 

Considering a wide variety of DM models, here we focus on ultralight ($m_\phi < 10 \un{eV}$) scalar field candidates characterized  by high mode occupation numbers ($\gg 1$); these can be described as classical fields.
The question of {\em microstructure} of DM is an open question~\cite{Berezinsky2014}. We simply 
split such fields into dichotomy of being either non-self-interacting or self-interacting. 
In the former case they are nearly uniformly distributed over the galaxies providing a uniform DM field background primarily oscillating at their Compton frequencies (``wavy" DM). Such candidates include pseudo-scalar axions and scalar dilatons/moduli. 
In the  case of self-interacting DM fields, of interest to our paper, self-interactions can lead to formation 
of clumps. Then DM can be viewed as a gas-like collection of gravitationally interacting clumps. Encounters with such objects may leave transient signals in measurement device data~\cite{DereviankoDM2014,AtomicReview2017}.
Examples of ``clumpy" DM models include Q-balls~\cite{Coleman1985,Lee1987,Kusenko2001,KimballQball2017},
Bose stars~\cite{Hogan1988,Barranco2013,Krippendorf2018},
topological defects \cite{Kibble1980,Vilenkin1985,Vilenkin1994},
axion quark nuggets~\cite{Zhitnitsky2019,Budker2019,Budker2020a}
and ``dark blobs"~\cite{Grabowska2018}.

A formation of DM clumps in the radiation era has been analyzed recently in Ref.~\cite{Brax2020}. The clump formation requires non-linear self-interactions of the scalar DM field. Non-linearities lead to cosmological fluid instability and the fluctuations of the scalar energy-density field lead to the formations of the clumps. Further, the clumps aggregate and afterwards follow the standard cold dark matter scenario. 
In this model, the gravitationally interacting clumps behave as the pre-requisite pressureless cosmological fluid. The scalar-field mass $m_\phi$ can span a wide range from $10^{-17} \un{eV}$ to $10 \un{GeV}$. 
The formed clumps span a wide range of scales and masses $M$, ranging from the size of atoms ($\sim$ angstroms) to that of galactic molecular clouds ($\sim$ parsec), and from a milligram to thousands of solar masses. For the considered range of parameters, the clumps do no collapse into black holes.
The clump mass-radius relation follows a power law, $M \sim R^n$, where the power $n=3,4,5$ depends on the details of the formation mechanisms and the self-interaction potential. Because of finite-size effects, these dark matter clumps are shown~\cite{Brax2020} to evade the microlensing constraints~\cite{Niikura2019}.

As we are interested in direct DM detection with laboratory instruments, local properties of DM are essential to interpreting such experiments. At the most basic level, our galaxy, the Milky Way, is embedded  into a DM halo and rotates through the halo. Astrophysical simulations  provide  estimates of DM properties in the Solar system (see, e.g., \cite{NesSal13}). The DM energy density in the vicinity of Solar system is estimated to be
 $\rho_\mathrm{DM} \approx 0.3  \un{GeV/cm^3}$, corresponding to $\sim$ one hydrogen atoms per three $\un{cm}^3$. Further, in the DM halo reference frame, the velocity distribution of DM objects is nearly Maxwellian with the dispersion of $v_\mathrm{vir} \sim 270\, \un{km/s}$ (referred to as the virial velocity in the literature) and a sharp cut-off at the galactic escape velocity $v_\mathrm{esc} \approx 650  \, \un{km/s}$.
Further, the Milky Way is a spiral galaxy  rotating through the DM halo. In particular, the Sun moves through the DM halo at galactic velocities $v_\mathrm{g} \approx 230\, \mathrm{km/s}$. For terrestrial experiments, there is an additional velocity modulation arising due to the Earth's orbital motion about the Sun, modulating the rate of encounters with DM objects. The period, phase, and amplitude of the modulation serve as unique DM signatures~\cite{Freese2013}.

A general challenge with searching for transient signals is that they are difficult to distinguish from conventional noise.
One approach~\cite{DereviankoDM2014,Pospelov2013}
is to use a network of devices, and search for the correlated propagation of transients that sweep through the network at galactic velocities, $v_g\sim300\un{km/s}$
(see also~\cite{GPSDM2018,Panelli2019,Budker2019,MasiaRoig2019,Jaeckel2020,Dailey2020}).
However, objects of spatial extent smaller than the network node separation would not produce such a signature.
Then one has to rely on unique signatures of the interactions with a single sensor that may differentiate them from the conventional noise.
Gravitational wave searches, for example, use both a correlated signal propagation across a network and a distinct signal pattern at each node~\cite{LIGO2016}.

If DM interacts with standard model particles, recurring encounters may cause perturbations in precision sensors.
If this were to lead only to a shift in the mean of the data it would be unobservable, as DM is always present.
Such interactions may, however, induce non-Gaussian signatures, such as an asymmetry in the data noise distribution, which are observable.
Further, we show that there would be an appreciable annual modulation in these signatures, that arises due to the Earth's orbital motion about the Sun, modulating the rate of encounters with DM objects.

Following these ideas, one may perform DM searches that are many orders of magnitude more sensitive than the existing constraints for certain models, and have discovery reach inaccessible by other means.
Our proposal is complimentary to  other ultralight DM searches, e.g.,~\cite{Arvanitaki2014,Tilburg2015,Hees2016,Wcislo2016,GPSDM2017,Kalaydzhyan2017,Wcislo-clock-network-2018,Savalle2020,RobertsTDs2020,Dailey2020}.
The technique proves particularly appealing for the parameter space of small clumps or high number density objects, where the expected encounter rate may be high.
Moreover, such searches may be performed using existing quantum sensors, making this an inexpensive avenue for potential discovery.
Finally, we note that while we focus on atomic clocks, the presented ideas  apply also to other precision instruments, such as magnetometers \cite{Pospelov2013,KimballQball2017}, interferometers \cite{StadnikLaser2015,StadnikLasInf2015,Arvanitaki2018},
gravimeters~\cite{Hu2019,McNally2019},
optical cavities~\cite{Savalle2020,Savalle2019},
and dipole moment searches~\cite{ACME2014,Budker2014,RobertsCosPRD2014,RobertsCosmic2014,Abel2017}.

\section{Results}

\subsection{Dark matter and atomic clocks}

\begin{figure}
\includegraphics[width=0.495\textwidth]{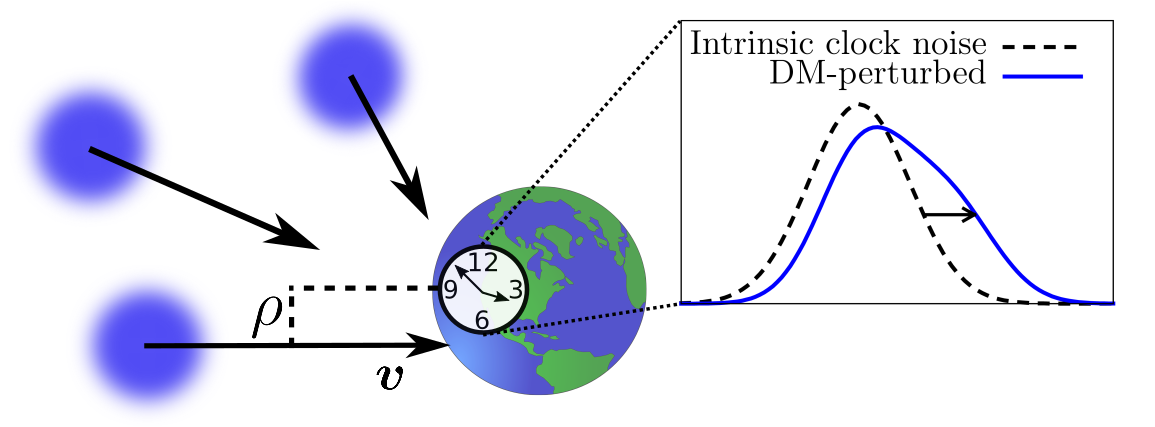}~
\caption{DM objects incident upon the Earth, and the induced shift and asymmetry in the clock noise distribution.}
\label{fig:mono}
\end{figure}

We consider interactions that lead to transient shifts in atomic transition frequencies of the form:
\be \label{eq:dw}
\delta\nu/{\nu_c} \propto \abs{\phi(\v{r},t)}^2,
\ee
where $\nu_c$ is the unperturbed frequency,
and $\phi$ is the DM field.
The proportionality constant depends on the DM model and the sensor.
As shown below, such interactions with macroscopic DM objects lead to an asymmetry in the noise distribution, as depicted in Fig.~\ref{fig:mono}.

The frequency excursion~(\ref{eq:dw}) leads to an additive term, $\chi$, in the time (phase) as measured by the clock:
\be \label{eq:s0}
\chi(t_j) = \int_{t_j-\tau_0}^{t_j}\frac{\delta\nu(\v{r},t)}{\nu_c}\d t,
\ee
where the phase differences (from one data sample to the next) are recorded for discrete values of elapsed time $t_j$.
Any DM encounter during the sampling interval will induce a shift in the measured phase.

Now we remark on some generic properties of  macroscopic DM objects.
We denote the radius of the objects as $R$, and the energy density inside each object as $\rho_{\phi}$.
By assuming the objects make up some fraction of total galactic DM density, these can be linked to $\CT$, the mean time between consecutive encounters of a given point-like instrument with a DM object as:
\be \label{eq:T}
{\CT}
 =  \frac{4\rho_{\phi}R}{3\rho_{\rm gal} v_g},
\ee
where $\rho_{\rm gal}$ is the total galactic energy density of the DM objects.
For simplicity, we assume such objects make up all of the dark matter, i.e., $ \rho_{\rm gal} = \rho_{\rm DM} \approx 0.4\un{GeV}\un{cm}^{-3}$~\cite{Bovy:2012tw}.
In a specific DM model, there may further be a model-dependent relations between $\rho_{\phi}$, $m_\phi$, and $R$; here we treat them as independent parameters.

To accumulate sufficient statistics, we require a high encounter rate ($\CT\ll1\,{\rm yr}$).
Then, Eq.~(\ref{eq:T}) leads to an upper bound on the mass of the objects.
For roughly Earth-sized objects, $R \sim R_E$, this is $M=\rho_{\phi} \frac{4}{3}\pi R^3/c^2\ll10^{-25}M_\sun$ ($M_\sun$ is the solar mass).
Bounds on massive DM objects from gravitational lensing constrain the mass to $M<10^{-16}M_\sun$ \cite{Hernandez2004,Gonzalez-Morales2013}
(see also Ref.~\cite{KimballQball2017}).
In this case, galactic structure formation would occur as per conventional cold dark matter theory~\cite{Blumenthal1984,Brax2020}.

\subsection{DM-induced variation of fundamental constants}

Now we specify the interactions of DM fields with the standard model.
The requirement that the $\phi$ sector retains the $U(1)$ symmetry naturally leads to portals quadratic in $\phi$ \cite{KimballQball2017}.
Those considered here can be expressed as
\begin{align}
\label{eq:Lquad}
\CL_{{\rm int},X}  & =\Gamma_{X} \, \phi\phi^{\ast} \mathcal{O}_{X},\\
\label{eq:Lderiv}
\CL_{{\rm int},X}' & =(\h c)^{2} \, \Gamma_{X}' \, (  \partial_{\mu}\phi) (  \partial^{\mu}\phi^{\ast}) \mathcal{O}_{X},
\end{align}
where $\mathcal{O}_{X}$ are various pieces of the standard model Lagrangian density,
$\mathcal{L}_{\mathrm{SM}}=\sum_{X}\mathcal{O}_{X}$.
The coupling constants $\Gamma_{X}$ and $\Gamma_{X}^{\prime}$ have units of
$[{\rm Energy}]^{-2}$ and $[{\rm Energy}]^{-4}$, respectively.

Both classes of portals lead to transient variation in the effective values of certain fundamental constants.
Those relevant to atomic clocks are the fine structure constant $\alpha$, the electron-proton mass ratio $m_e/m_p$, and the ratio of the light quark mass to the QCD energy scale $m_q/\Lambda_{\rm QCD}$.
For concreteness, we focus on the quadratic portal (\ref{eq:Lquad}); we will generalize the discussion to the derivative portal~(\ref{eq:Lderiv}) in Sec.~\ref{Sec:Discussion}.
Generically, for each such constant $X$, we may express its fractional variation (inside the DM object) as
\begin{align}
\frac{\delta X}{X}  & =\Gamma_{X} \abs{\phi}^{2} =\Gamma_{X} {\phi_0}^{2},  \label{eq:lintQ}
\end{align}
where $|{\phi_0}|$ is the maximum of the field amplitude inside the DM object.
In general this is model-dependent; e.g., for topological defects $R\simeq\h/(m_\phi c)$, which coupled with Eq.~(\ref{eq:T}), leads to
$|{\phi_0}|^2 = \h c \rho_{DM}v_g\CT R$~\cite{DereviankoDM2014}.
Such DM-induced variations in fundamental constants lead to transient shifts in atomic transition frequencies:
\be \label{eq:Geff}
\frac{\delta\nu(t)}{\nu_c}
= \sum_X K_X\frac{\delta X(t)}{X}
= \Gamma_{\rm eff} \, \phi(t)^2.
\ee
Here, $\Gamma_{\rm eff}\equiv\sum_X K_X\Gamma_X$, and $K_X$ are sensitivity coefficients that quantify the response of the atomic transition to the variation in a given fundamental constant \cite{Angstmann2004,Dinh2009}.
Eq.~(\ref{eq:Geff}) establishes the proportionality factor in Eq.~(\ref{eq:dw}).

\subsection{DM-induced asymmetry and skewness}

Now we consider the statistics and observable effects of DM encounters with atomic clocks.
Not every encounter  imparts the same signal magnitude, as the DM velocities and impact parameters differ.
However, for the considered couplings 
the {\em sign} of the perturbation remains the same, since it is set only by the sign of $\Gamma_{\rm eff}$ (\ref{eq:Geff}).
This leads to an asymmetry in the observed data noise distribution.
It may be possible to observe this asymmetry, even if individual events cannot be resolved or the perturbations are well below the noise.

The observed clock noise value at a given time is $s = \eta + \chi$ if there was a DM interaction during the sampling interval, and $s = \eta$ otherwise.
Here, $\eta$ is the conventional physics noise.
If $p_\chi$ is the distribution for induced DM signals (in the absence of noise), the observed probability distribution for clock excursions reads
\be \label{eq:ps}
p_s(s) =
	\frac{\t_0}{\CT} \int_{-\infty}^\infty p_\eta(\eta)p_\chi(s-\eta)\,\d\eta
	+
	(1-\frac{\t_0}{\CT}) \,p_\eta(s),
\ee
where
$p_\eta$ is the intrinsic noise distribution, and $\t_0$ is the data sampling interval (averaging time).
For $p_\eta$, we assume Gaussian noise with standard deviation $\sigma$.
Formally, this is the assumption of white frequency noise, which is typically dominant for atomic clocks.
For clocks, $\sigma$ is related to the Allan deviation as $\sigma \approx \t_{0} \sigma_{y}(\t_{0})$.
While other noise processes affect the clocks, we assume that $p_\eta$ is symmetric.
Even if it were not the case, the annual modulation discussed below would  remain an observable DM signature.

The skewness,  defined as the third standard moment,
\be
\k_3 \equiv \frac{\braket{(x-\bar x)^3}}{\braket{(x-\bar x)^2}^{3/2}},
\ee
is a measure of the asymmetry in the distribution for random variable $x$.
The uncertainty in the sample skewness is
$\delta\k_3 = \sqrt{{6}/{N}}$, where $N$ is the number of data points.
The expected value of the DM-induced skewness can be calculated for a given model as
\be \label{eq:DMskew}
\k_3 = \frac{1}{\s_s^3}\int_{-\infty}^\infty s^3 p_s(s+\bar s) \,\d s,
\ee
where the mean $\bar s$ and variance $\s_s^2$ are from $p_s$ (\ref{eq:ps}).
In addition to $\k_3$, there are DM-induced contributions to other moments, such as  kurtosis and variance.

To compute the expected DM-induced skewness, we
first determine the DM signal distribution, $p_\chi$.
The magnitude of each DM signal depends on the velocity, $v$, and impact parameter, $\rho$.
We take the $v$ distribution, $f_{v}$, to be that of  the standard halo model (see, e.g., Ref.~\cite{Freese2013}).
The $\rho$ distribution comes from geometric arguments: for ball-like (spherical) objects it is $p_\rho(\rho) = {2\rho}/{R^2}$.

For objects small enough that they traverse the clock within one sampling interval, i.e., $R < v\tau_0$, the DM signal per encounter contributes to just a single data point, and has magnitude:
\be \label{eq:QballChi}
\chi = \chi_0  \frac{v_g}{v} \, \sqrt{1-\rho^2/R^2}
\ee
for $\rho<R$ ($\chi=0$ otherwise),
where $\chi_0 \equiv \Gamma_{\rm eff} \phi_0^2 {R}/{v_g}$.
Without loss of generality, we take $\chi_0>0$ from here on.

While it is not required for the further analysis, to connect with the particle physics DM searches, it is instructive to introduce a cross-section $\sigma_\chi$ which has a meaning of  accumulation rate of normalized (unit-less) DM signal $\chi/\tau_0$ due to interaction with a spatially uniform beam of DM blobs of velocity $v$. This involves averaging $\chi$, Eq.~(\ref{eq:QballChi}), over impact parameters with probability $p_\rho(\rho)$,
\be
\sigma_\chi = \frac{2}{3} \frac{\chi_0}{\tau_0} \frac{v_g}{v} \pi R^2 \,.
\label{Eq:x-section}
\ee
The cross-section is inversely proportional to velocity, reflecting the fact that the longer the DM blob bulk overlaps with the sensor, the larger the DM-induced frequency excursion~(\ref{eq:s0}) is.

Combining Eq.~(\ref{eq:QballChi}) with the $\rho$ and $v$ probability distributions, the signal magnitude distribution is
\be \label{eq:pChi}
p_\chi(\chi)
= \frac{2\chi}{\chi_0^2 \, v_g^2} \int_0^{\frac{v_g\chi_0}{\chi}}v^2 f_v(v)\,\d v
\approx  \frac{2\chi}{\chi_0^2}.
\ee
In order to extract simple analytic results we made an approximation here, noting that $f_v$ peaks at $v_g$; we have confirmed the adequacy of this simplification numerically~\cite{code}.
We have also verified numerically that the approximate result in Eq.~(\ref{eq:pChi}) also holds adequately for other DM object profiles, such as Gaussian monopoles.

From the above, the DM-induced skewness can be found analytically (to leading order in $\tau_0/\CT$):
\be \label{eq:DMk3}
\k_3 \approx \frac{2\tau_0 \chi_0^3}{5\CT\s^3}.
\ee
Requiring that $\k_3>\delta\k_3$, and noting that the number of measurements  $N=T_{\rm obs}/\tau_0$, where $T_{\rm obs}$ is the total observation time, implies the smallest detectable signal satisfies

\be \label{eq:x03r0}
\frac{\abs{\chi_0}^{3}}{\CT}  \gtrsim \frac{5\s^3}{2} \sqrt{\frac{6}{T_{\rm obs}\tau_0}}.
\ee
This formula is assuming that the uncertainty in the observed skewness is given by the statistical sample uncertainty, $\delta\k_3 = \sqrt{{6}/{N}}$.
This is a reasonable assumption, though in actual experiments, the true uncertainty should be estimated (e.g., by calculating the skewness for multiple randomised subsets of the data).
For the general case, if the maximum observed skewness is constrained to be below $\kappa_3^{\rm max}$, then constraints on the combination of parameters may be placed:
\be\label{eq:general-constraint}
\Gamma_{\rm eff}\,|\phi_0|^2 < \frac{\sigma v_g}{R}\left[ (5/2)(\CT/\t)\kappa_3^{\rm max}\right]^{1/3}.
\ee
The form of $\phi_0$ (the field amplitude inside the DM object) is model-dependent; a few specific examples will be considered below.

\subsection{Symmetric non-Gaussian signatures}

As well as the skewness, other non-Gaussian signatures will also be induced in the precision device noise due to interactions with dark matter.
This is important, for example, in situations where the frequency deviation (\ref{eq:dw}) may occur with either sign (this may occur in some dark matter models, for example, for linear rather than quadratic couplings).
In such cases, no asymmetric moments are induced, though there are still symmetric non-Gaussian DM-induced signatures.
In particular, there is a DM contribution to the variance and to the kurtosis, the fourth standard moment defined 
\be
\k_4 \equiv \frac{\braket{(x-\bar x)^4}}{\braket{(x-\bar x)^2}^{2}}-3.
\ee
Respectively, these are
\begin{align}
\Delta\s^2 &\approx \frac{R_0\tau_0\chi_0^2}{2}
\\
\kappa_4 &\approx \frac{R_0\tau_0\chi_0^4}{3\sigma^4}.
\end{align}

Of course, symmetric non-Gaussianities are difficult to distinguish from regular noise, and the average DM contribution to the variance is entirely unobservable.
However, due to the galactic motion of the Earth, annual modulations in these signatures, as well as the skewness, are induced, which are observable.

\subsection{Annual modulation}

As the Earth orbits the Sun, there is an annual modulation in the addition of their velocities.
This causes an annual modulation in the Earth's velocity relative to the galactic DM halo, and hence to the mean DM encounter rate.
We may therefore express the rate, $\CR=1/\CT$, as

\begin{equation}
{\CR}(t)={\CR}^{(0)}\left(1 + \frac{\Delta v}{v_g}\right)\cos(\Omega t + \varphi),
\end{equation}

where  $\Omega=2\pi/{\rm yr}$,  $\varphi$ is the phase with $\Omega t + \varphi=0$ on 2 June, and
${\Delta v}/{v_g}\approx0.05$~\cite{Freese2013}.

\begin{figure}[ht]
   \centering
	\includegraphics[width=0.475\textwidth]{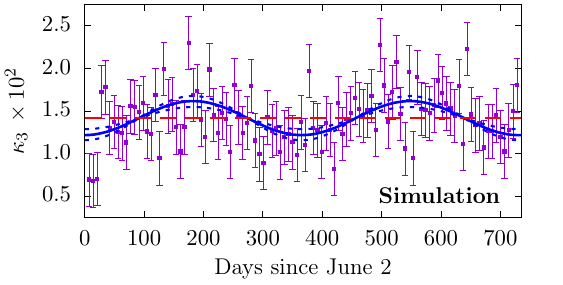}
	\caption{Simulation for two years of data ($\t_0=1\,$s), with DM signals ($R_0\t_0=0.01$, $\chi_0/\s=1$) including the annual velocity modulation~\cite{code}.
The skewness is calculated for each week of data (purple squares, with $\sqrt{{6}/{N}}$ error bars).
The extracted modulation amplitude is $\k_3^{(m)}=0.2\times10^{-2}$ (\ref{eq:k3m}).
The solid blue curve is the best-fit cosine, and the dotted lines are the uncertainties. The dashed red curve shows the mean $\k_3$.
}
\label{fig:modulation}
\end{figure}

Then, the skewness
(and other moments) becomes time-dependent:
\be \label{eq:k3t}
\k_3(t) \approx \k_3^{(0)} - \k_3^{(m)} \cos(\Omega t + \varphi).
\ee
The DM-induced skewness (\ref{eq:DMk3}) scales linearly with the rate, and as the cube of the mean signal magnitude.
The mean signal magnitude scales inversely with velocity (\ref{eq:QballChi}).
Therefore, the modulation amplitude is
\be
\label{eq:k3mod}
\k_3^{(m)} = 2\frac{\Delta v}{v_0}\k_3 \sim10\%.
\ee
We demonstrate this using simulated data in Fig.~\ref{fig:modulation}.
Similarly, the annual modulation in the kurtosis is
\[
\k_4^{(m)} = 3\frac{\Delta v}{v_0}\k_3 \sim15\%.
\]

If the data is divided into $M$ time bins, each consisting of $N_M=N/M$ points, with the skewness calculated for each bin,
the modulation amplitude can be extracted as
\be\label{eq:k3m}
 \k_3^{(m)} = 2 \frac{ \abs{\widetilde \k_3(1/{\rm yr})}}{M}  \pm \delta \k_3^{(m)} ,
\ee
where
$\widetilde \k_3$ is the Fourier transform of  $\k_3(t)$.
The sample uncertainty,
$\delta \k_3^{(m)}  \approx 2 \sqrt{{6}/{N}},$
is independent of the number of bins.
However, the requirement to have several encounters per bin limits the sensitivity region to $\CT\ll N_M\tau_0 = T_{\rm obs}/M$.

To detect the annual modulation in the skewness, we require that $\k_3^{(m)}>\delta \k^{(m)}_3$.
This implies that we require signals with combination $\chi_0^3/\CT$ that are larger by a factor ${v}/{\Delta v}\approx20$ compared to the result for the mean skewness (\ref{eq:x03r0}).
Or, for a fixed value of ${\CT}$, signals that are $\sim$\,3 times larger.
Nevertheless, it is important that there are signatures unique to DM (namely, the modulation phase, period, and amplitude) that can be sought in such experiments.
If a skewness is present in the data, one may exclude DM origins if the modulation is absent.

\section{Discussion}
\label{Sec:Discussion}

As an illustrative example, we analyze six years of archival atomic clock data \cite{JPLigsac,MurphyJPL2015} from the comparison of several Cs GPS satellite clocks to an Earth-based H-maser.
We use the same GPS data used by us in Ref.~\cite{GPSDM2017}; see Refs.~\cite{GPSDM2017,Panelli2019} for a description of the GPS clock data relevant to the analysis.
The calculated skewness in the clock-comparison residuals  is
\be
\k_3({\rm Cs}) = (0.1\pm47.0)\times10^{-3},
\ee
which, at the 68\% confidence level, implies
$\abs{\k_3}<4.7\times10^{-2}$
(for this GPS data, $\sigma\simeq0.09\un{ns}$, and $\t_0=30\un{s}$ \cite{GPSDM2018}).
The uncertainty in $\kappa_3$ was found by calculating the skewness for each day of data separately; note that this is larger than the assumed sample skewness due to the presence of non-Gaussian noise (including outliers, which are not removed) in the data. 
From Eq.~(\ref{eq:general-constraint}), we can thus place constraints on the $\Gamma_X$ couplings.
Importantly, this allows one to place constraints on  couplings for macroscopic DM objects with radii $R<10^4\un{km}$, the region that is otherwise inaccessible using global network methods \cite{GPSDM2018}.

To demonstrate this in more concrete terms, we assume here a scalar field DM model for which the energy density inside the DM objects scales as $\rho_{\phi}\sim\phi_0^2 m_\phi^2$, and the size of the objects is set by the Compton wavelength $R\sim \hbar/m_\phi c$.
This is consistent, for example, with topological defect models~\cite{DereviankoDM2014} (we note however, that this is just an example, and for other models, different relations will hold).
In this case, if no signal is observed, the model may be constrained as
\be
|\Gamma_{\rm eff}| R^2 < \frac{\sigma |\k_3^{\rm max}|^{1/3}}{\hbar c \, \rho_{\rm DM} \mathcal{T}^{2/3}\tau_0^{1/3}}. \label{Eq:Constraint}
\ee

\begin{figure}
   \centering
	\includegraphics[width=0.475\textwidth]{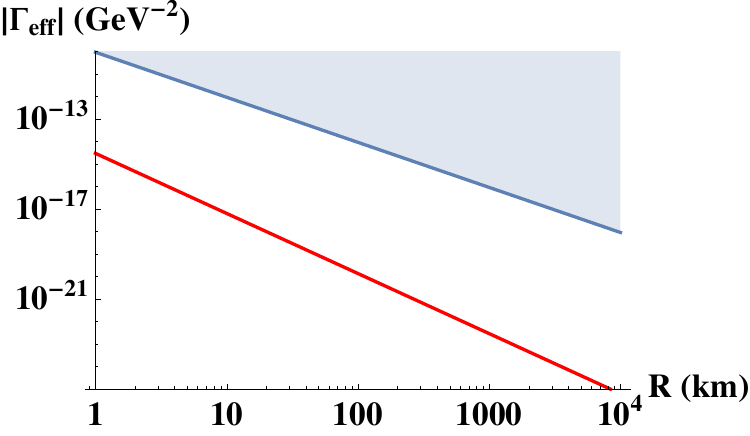}
\caption{Preliminary results: constraints on a general scalar ``DM blob" model with quadratic interactions as per Eq.~(\ref{eq:Geff}), for the average time between encounter $\CT=1\,$day.
The shaded blue region is the preliminary constraints found in this work from six years of data from the Cs GPS clocks.
The red line shows the potential discovery reach for one year of data from a single laboratory optical Sr clock as described in the text.  These constraints can be easily rescaled into those for a derivative portal coupling strengths via Eq.~(\protect\ref{Eq:RescalingQuad2Derivative}). 
}
\label{fig:prelim}
\end{figure}

Preliminary results for such a model from the above analysis of the Cs GPS clocks is presented in Fig.~\ref{fig:prelim}.
Note that results from the experiments in Refs.~\cite{GPSDM2017,RobertsTDs2020,Wcislo2016,Wciso2018} do not apply in the considered parameter range.
Also shown is the projected sensitivity for 1 year of data from an optical Sr clock, assuming $\sigma\sim10^{-16}\un{s}$ at averaging time of $\t_0=1\un{s}$.
Such clocks have been used recently for DM searches, both for ``clumpy" and oscillating DM models, in  Refs.~\cite{RobertsTDs2020,Wciso2018}; details of the clock performance are given in those works (see also discussion of clock servo loop and averaging times relevant to DM searches in Refs.~\cite{GPSDM2017,RobertsTDs2020}).
This projection takes into account that the optimal averaging time to use when searching for DM objects of radius $R$ is $\tau_{\rm avg}\simeq R/v_g$.

The results of our analysis for the quadratic portal (\ref{eq:Lquad}) can be easily translated into the constraints on the derivative portal (\ref{eq:Lderiv}) by noticing that $(  \partial_{\mu}\phi) (  \partial^{\mu}\phi^{\ast}) \approx - |\bm{\nabla} \phi|^2$, where we neglected the time derivative because of the non-relativistic nature of cold DM. Further, for a Gaussian-profiled ``blob''  $|\bm{\nabla} \phi|^2 \sim \phi_0^2/R^2$. Thereby, 
\be
\Gamma_{X}' \sim - \Gamma_{X} R^2/(\h c)^{2} \label{Eq:RescalingQuad2Derivative}
\ee
and the constraint~(\ref{Eq:Constraint})
translates into 
\be
|\Gamma_{\rm eff}'| < \frac{\sigma |\k_3^{\rm max}|^{1/3}}{(\hbar c)^3 \, \rho_{\rm DM} \mathcal{T}^{2/3}\tau_0^{1/3}}. \label{Eq:Constraint}
\ee

Can our DM observable, the noise asymmetry,  be mimicked by fluctuations in DM energy density, $\rho_\mathrm{DM}$? It can not. Indeed, the sign of the frequency perturbation~(\ref{eq:Geff}) due to a single 
DM blob is fixed. DM energy density (or the number density of DM blobs) affects the encounter rate of DM blobs with the sensor. However, since the sign of the DM-induced perturbation remains the same, all individual perturbations add coherently. If DM energy density fluctuates, it would only scale the DM blob flux and thus the observable. 

Another relevant point recently raised in the  literature~\cite{StochasticAmplitude2019} is the effect of DM energy density fluctuations on the coupling strength constraints. For scalar fields, the effective sensitivity was shown to be reduced by a factor of a few. Considering the logarithmic scale of Fig.~\ref{fig:prelim} and the preliminary, illustrative nature of our results, this corrective factor would not affect our conclusions.

\section{Conclusion}

In this work, we proposed a new dark matter signature: an asymmetry (and other non-Gaussianities) that may be induced in the noise distributions of precision quantum sensors, such as atomic clocks.
Such signatures may be induced by dark matter candidates composed of self-interacting ultralight quantum fields that form macroscopic objects, examples of which include Q-balls and topological defects.
Further, we demonstrate that there would be a sizeable annual modulation in these signatures due to the annual variation of the Earth velocity with respect to dark matter halo.
As an application of our formalism, we use 6 years of data from the atomic clocks on board GPS satellites to place constraints on a scalar dark matter model, and show projections for future experiments based on laboratory clocks.
This technique allows one to search for DM models that would otherwise be undetectable using existing experiments.

\bibliography{library-apd,asym,code}

\begin{thebibliography}{67}%
\makeatletter
\providecommand \@ifxundefined [1]{%
 \@ifx{#1\undefined}
}%
\providecommand \@ifnum [1]{%
 \ifnum #1\expandafter \@firstoftwo
 \else \expandafter \@secondoftwo
 \fi
}%
\providecommand \@ifx [1]{%
 \ifx #1\expandafter \@firstoftwo
 \else \expandafter \@secondoftwo
 \fi
}%
\providecommand \natexlab [1]{#1}%
\providecommand \enquote  [1]{``#1''}%
\providecommand \bibnamefont  [1]{#1}%
\providecommand \bibfnamefont [1]{#1}%
\providecommand \citenamefont [1]{#1}%
\providecommand \href@noop [0]{\@secondoftwo}%
\providecommand \href [0]{\begingroup \@sanitize@url \@href}%
\providecommand \@href[1]{\@@startlink{#1}\@@href}%
\providecommand \@@href[1]{\endgroup#1\@@endlink}%
\providecommand \@sanitize@url [0]{\catcode `\\12\catcode `\$12\catcode
  `\&12\catcode `\#12\catcode `\^12\catcode `\_12\catcode `\%12\relax}%
\providecommand \@@startlink[1]{}%
\providecommand \@@endlink[0]{}%
\providecommand \url  [0]{\begingroup\@sanitize@url \@url }%
\providecommand \@url [1]{\endgroup\@href {#1}{\urlprefix }}%
\providecommand \urlprefix  [0]{URL }%
\providecommand \Eprint [0]{\href }%
\providecommand \doibase [0]{http://dx.doi.org/}%
\providecommand \selectlanguage [0]{\@gobble}%
\providecommand \bibinfo  [0]{\@secondoftwo}%
\providecommand \bibfield  [0]{\@secondoftwo}%
\providecommand \translation [1]{[#1]}%
\providecommand \BibitemOpen [0]{}%
\providecommand \bibitemStop [0]{}%
\providecommand \bibitemNoStop [0]{.\EOS\space}%
\providecommand \EOS [0]{\spacefactor3000\relax}%
\providecommand \BibitemShut  [1]{\csname bibitem#1\endcsname}%
\let\auto@bib@innerbib\@empty
\bibitem [{\citenamefont {Weinberg}(2008)}]{weinberg-cosmology-2008}%
  \BibitemOpen
  \bibfield  {author} {\bibinfo {author} {\bibfnamefont {S.}~\bibnamefont
  {Weinberg}},\ }\href {https://books.google.com/books?id=nqQZdg020fsC} {\emph
  {\bibinfo {title} {{Cosmology}}}},\ Cosmology\ (\bibinfo  {publisher} {Oxford
  University Press},\ \bibinfo {address} {New York, NY, USA},\ \bibinfo {year}
  {2008})\BibitemShut {NoStop}%
\bibitem [{\citenamefont {Peebles}(2003)}]{Peebles2003}%
  \BibitemOpen
  \bibfield  {author} {\bibinfo {author} {\bibfnamefont {P.~J.~E.}\
  \bibnamefont {Peebles}},\ }\href@noop {} {\bibfield  {journal} {\bibinfo
  {journal} {Rev.\ Mod.\ Phys.}\ }\textbf {\bibinfo {volume} {75}},\ \bibinfo
  {pages} {559} (\bibinfo {year} {2003})}\BibitemShut {NoStop}%
\bibitem [{\citenamefont {Bertone}\ \emph {et~al.}(2005)\citenamefont
  {Bertone}, \citenamefont {Hooper},\ and\ \citenamefont {Silk}}]{Bertone2005}%
  \BibitemOpen
  \bibfield  {author} {\bibinfo {author} {\bibfnamefont {G.}~\bibnamefont
  {Bertone}}, \bibinfo {author} {\bibfnamefont {D.}~\bibnamefont {Hooper}}, \
  and\ \bibinfo {author} {\bibfnamefont {J.}~\bibnamefont {Silk}},\ }\href
  {\doibase 10.1016/j.physrep.2004.08.031} {\bibfield  {journal} {\bibinfo
  {journal} {Phys. Rep.}\ }\textbf {\bibinfo {volume} {405}},\ \bibinfo {pages}
  {279} (\bibinfo {year} {2005})}\BibitemShut {NoStop}%
\bibitem [{\citenamefont {Feng}(2010)}]{Feng2010}%
  \BibitemOpen
  \bibfield  {author} {\bibinfo {author} {\bibfnamefont {J.~L.}\ \bibnamefont
  {Feng}},\ }\href {\doibase 10.1146/annurev-astro-082708-101659} {\bibfield
  {journal} {\bibinfo  {journal} {Ann. Rev. Astro. Astrophys.}\ }\textbf
  {\bibinfo {volume} {48}},\ \bibinfo {pages} {495} (\bibinfo {year}
  {2010})}\BibitemShut {NoStop}%
\bibitem [{\citenamefont {Matarrese}\ \emph {et~al.}(2011)\citenamefont
  {Matarrese}, \citenamefont {Colpi}, \citenamefont {Gorini},\ and\
  \citenamefont {Moschella}}]{Matarrese2011}%
  \BibitemOpen
  \bibfield  {author} {\bibinfo {author} {\bibfnamefont {S.}~\bibnamefont
  {Matarrese}}, \bibinfo {author} {\bibfnamefont {M.}~\bibnamefont {Colpi}},
  \bibinfo {author} {\bibfnamefont {V.}~\bibnamefont {Gorini}}, \ and\ \bibinfo
  {author} {\bibfnamefont {U.}~\bibnamefont {Moschella}},\ }\href@noop {}
  {\emph {\bibinfo {title} {{Dark Matter and Dark Energy: A Challenge for
  Modern Cosmology}}}}\ (\bibinfo  {publisher} {Springer},\ \bibinfo {address}
  {Berlin Heidelberg},\ \bibinfo {year} {2011})\BibitemShut {NoStop}%
\bibitem [{\citenamefont {Blumenthal}\ \emph {et~al.}(1984)\citenamefont
  {Blumenthal}, \citenamefont {Faber}, \citenamefont {Primack},\ and\
  \citenamefont {Rees}}]{Blumenthal1984}%
  \BibitemOpen
  \bibfield  {author} {\bibinfo {author} {\bibfnamefont {G.~R.}\ \bibnamefont
  {Blumenthal}}, \bibinfo {author} {\bibfnamefont {S.~M.}\ \bibnamefont
  {Faber}}, \bibinfo {author} {\bibfnamefont {J.~R.}\ \bibnamefont {Primack}},
  \ and\ \bibinfo {author} {\bibfnamefont {M.~J.}\ \bibnamefont {Rees}},\
  }\href {\doibase 10.1038/311517a0} {\bibfield  {journal} {\bibinfo  {journal}
  {Nature}\ }\textbf {\bibinfo {volume} {311}},\ \bibinfo {pages} {517}
  (\bibinfo {year} {1984})}\BibitemShut {NoStop}%
\bibitem [{\citenamefont {{The XENON Collaboration}}(2017)}]{XENON-1T-2017}%
  \BibitemOpen
  \bibfield  {author} {\bibinfo {author} {\bibnamefont {{The XENON
  Collaboration}}},\ }\href {\doibase 10.1103/PhysRevLett.119.181301}
  {\bibfield  {journal} {\bibinfo  {journal} {Phys. Rev. Lett.}\ }\textbf
  {\bibinfo {volume} {119}},\ \bibinfo {pages} {181301} (\bibinfo {year}
  {2017})}\BibitemShut {NoStop}%
\bibitem [{\citenamefont {Liu}\ \emph {et~al.}(2017)\citenamefont {Liu},
  \citenamefont {Chen},\ and\ \citenamefont {Ji}}]{Liu2017}%
  \BibitemOpen
  \bibfield  {author} {\bibinfo {author} {\bibfnamefont {J.}~\bibnamefont
  {Liu}}, \bibinfo {author} {\bibfnamefont {X.}~\bibnamefont {Chen}}, \ and\
  \bibinfo {author} {\bibfnamefont {X.}~\bibnamefont {Ji}},\ }\href {\doibase
  10.1038/nphys4039} {\bibfield  {journal} {\bibinfo  {journal} {Nat. Phys.}\
  }\textbf {\bibinfo {volume} {13}},\ \bibinfo {pages} {212} (\bibinfo {year}
  {2017})}\BibitemShut {NoStop}%
\bibitem [{\citenamefont {Bertone}\ and\ \citenamefont
  {Tait}(2018)}]{Bertone2018}%
  \BibitemOpen
  \bibfield  {author} {\bibinfo {author} {\bibfnamefont {G.}~\bibnamefont
  {Bertone}}\ and\ \bibinfo {author} {\bibfnamefont {T.~M.~P.}\ \bibnamefont
  {Tait}},\ }\href {\doibase 10.1038/s41586-018-0542-z} {\bibfield  {journal}
  {\bibinfo  {journal} {Nature}\ }\textbf {\bibinfo {volume} {562}},\ \bibinfo
  {pages} {51} (\bibinfo {year} {2018})},\ \Eprint
  {http://arxiv.org/abs/1810.01668} {arXiv:1810.01668} \BibitemShut {NoStop}%
\bibitem [{\citenamefont {Berezinsky}\ \emph {et~al.}(2014)\citenamefont
  {Berezinsky}, \citenamefont {Dokuchaev},\ and\ \citenamefont
  {Eroshenko}}]{Berezinsky2014}%
  \BibitemOpen
  \bibfield  {author} {\bibinfo {author} {\bibfnamefont {V.~S.}\ \bibnamefont
  {Berezinsky}}, \bibinfo {author} {\bibfnamefont {V.~I.}\ \bibnamefont
  {Dokuchaev}}, \ and\ \bibinfo {author} {\bibfnamefont {Y.~N.}\ \bibnamefont
  {Eroshenko}},\ }\href {\doibase 10.3367/UFNe.0184.201401a.0003} {\bibfield
  {journal} {\bibinfo  {journal} {Physics-Uspekhi}\ }\textbf {\bibinfo {volume}
  {57}},\ \bibinfo {pages} {1} (\bibinfo {year} {2014})}\BibitemShut {NoStop}%
\bibitem [{\citenamefont {Derevianko}\ and\ \citenamefont
  {Pospelov}(2014)}]{DereviankoDM2014}%
  \BibitemOpen
  \bibfield  {author} {\bibinfo {author} {\bibfnamefont {A.}~\bibnamefont
  {Derevianko}}\ and\ \bibinfo {author} {\bibfnamefont {M.}~\bibnamefont
  {Pospelov}},\ }\href {\doibase 10.1038/nphys3137} {\bibfield  {journal}
  {\bibinfo  {journal} {Nat. Phys.}\ }\textbf {\bibinfo {volume} {10}},\
  \bibinfo {pages} {933} (\bibinfo {year} {2014})}\BibitemShut {NoStop}%
\bibitem [{\citenamefont {Safronova}\ \emph {et~al.}(2018)\citenamefont
  {Safronova}, \citenamefont {Budker}, \citenamefont {DeMille}, \citenamefont
  {Kimball}, \citenamefont {Derevianko},\ and\ \citenamefont
  {Clark}}]{AtomicReview2017}%
  \BibitemOpen
  \bibfield  {author} {\bibinfo {author} {\bibfnamefont {M.~S.}\ \bibnamefont
  {Safronova}}, \bibinfo {author} {\bibfnamefont {D.}~\bibnamefont {Budker}},
  \bibinfo {author} {\bibfnamefont {D.}~\bibnamefont {DeMille}}, \bibinfo
  {author} {\bibfnamefont {D.~F.~J.}\ \bibnamefont {Kimball}}, \bibinfo
  {author} {\bibfnamefont {A.}~\bibnamefont {Derevianko}}, \ and\ \bibinfo
  {author} {\bibfnamefont {C.~W.}\ \bibnamefont {Clark}},\ }\href {\doibase
  10.1103/RevModPhys.90.025008} {\bibfield  {journal} {\bibinfo  {journal}
  {Rev. Mod. Phys.}\ }\textbf {\bibinfo {volume} {90}},\ \bibinfo {pages}
  {025008} (\bibinfo {year} {2018})}\BibitemShut {NoStop}%
\bibitem [{\citenamefont {Coleman}(1985)}]{Coleman1985}%
  \BibitemOpen
  \bibfield  {author} {\bibinfo {author} {\bibfnamefont {S.}~\bibnamefont
  {Coleman}},\ }\href {\doibase 10.1016/0550-3213(85)90286-X} {\bibfield
  {journal} {\bibinfo  {journal} {Nucl. Phys. B}\ }\textbf {\bibinfo {volume}
  {262}},\ \bibinfo {pages} {263} (\bibinfo {year} {1985})}\BibitemShut
  {NoStop}%
\bibitem [{\citenamefont {Lee}(1987)}]{Lee1987}%
  \BibitemOpen
  \bibfield  {author} {\bibinfo {author} {\bibfnamefont {T.~D.}\ \bibnamefont
  {Lee}},\ }\href {\doibase 10.1103/PhysRevD.35.3637} {\bibfield  {journal}
  {\bibinfo  {journal} {Phys. Rev. D}\ }\textbf {\bibinfo {volume} {35}},\
  \bibinfo {pages} {3637} (\bibinfo {year} {1987})}\BibitemShut {NoStop}%
\bibitem [{\citenamefont {Kusenko}\ and\ \citenamefont
  {Steinhardt}(2001)}]{Kusenko2001}%
  \BibitemOpen
  \bibfield  {author} {\bibinfo {author} {\bibfnamefont {A.}~\bibnamefont
  {Kusenko}}\ and\ \bibinfo {author} {\bibfnamefont {P.~J.}\ \bibnamefont
  {Steinhardt}},\ }\href {\doibase 10.1103/PhysRevLett.87.141301} {\bibfield
  {journal} {\bibinfo  {journal} {Phys. Rev. Lett.}\ }\textbf {\bibinfo
  {volume} {87}},\ \bibinfo {pages} {141301} (\bibinfo {year}
  {2001})}\BibitemShut {NoStop}%
\bibitem [{\citenamefont {Kimball}\ \emph {et~al.}(2018)\citenamefont
  {Kimball}, \citenamefont {Budker}, \citenamefont {Eby}, \citenamefont
  {Pospelov}, \citenamefont {Pustelny}, \citenamefont {Scholtes}, \citenamefont
  {Stadnik}, \citenamefont {Weis},\ and\ \citenamefont
  {Wickenbrock}}]{KimballQball2017}%
  \BibitemOpen
  \bibfield  {author} {\bibinfo {author} {\bibfnamefont {D.~F.~J.}\
  \bibnamefont {Kimball}}, \bibinfo {author} {\bibfnamefont {D.}~\bibnamefont
  {Budker}}, \bibinfo {author} {\bibfnamefont {J.}~\bibnamefont {Eby}},
  \bibinfo {author} {\bibfnamefont {M.}~\bibnamefont {Pospelov}}, \bibinfo
  {author} {\bibfnamefont {S.}~\bibnamefont {Pustelny}}, \bibinfo {author}
  {\bibfnamefont {T.}~\bibnamefont {Scholtes}}, \bibinfo {author}
  {\bibfnamefont {Y.~V.}\ \bibnamefont {Stadnik}}, \bibinfo {author}
  {\bibfnamefont {A.}~\bibnamefont {Weis}}, \ and\ \bibinfo {author}
  {\bibfnamefont {A.}~\bibnamefont {Wickenbrock}},\ }\href {\doibase
  10.1103/PhysRevD.97.043002} {\bibfield  {journal} {\bibinfo  {journal} {Phys.
  Rev. D}\ }\textbf {\bibinfo {volume} {97}},\ \bibinfo {pages} {043002}
  (\bibinfo {year} {2018})}\BibitemShut {NoStop}%
\bibitem [{\citenamefont {Hogan}\ and\ \citenamefont {Rees}(1988)}]{Hogan1988}%
  \BibitemOpen
  \bibfield  {author} {\bibinfo {author} {\bibfnamefont {C.~J.}\ \bibnamefont
  {Hogan}}\ and\ \bibinfo {author} {\bibfnamefont {M.~J.}\ \bibnamefont
  {Rees}},\ }\href {\doibase 10.1016/0370-2693(88)91655-3} {\bibfield
  {journal} {\bibinfo  {journal} {Phys. Lett. B}\ }\textbf {\bibinfo {volume}
  {205}},\ \bibinfo {pages} {228} (\bibinfo {year} {1988})}\BibitemShut
  {NoStop}%
\bibitem [{\citenamefont {Barranco}\ \emph {et~al.}(2013)\citenamefont
  {Barranco}, \citenamefont {Monteverde},\ and\ \citenamefont
  {Delepine}}]{Barranco2013}%
  \BibitemOpen
  \bibfield  {author} {\bibinfo {author} {\bibfnamefont {J.}~\bibnamefont
  {Barranco}}, \bibinfo {author} {\bibfnamefont {A.~C.}\ \bibnamefont
  {Monteverde}}, \ and\ \bibinfo {author} {\bibfnamefont {D.}~\bibnamefont
  {Delepine}},\ }\href {\doibase 10.1103/PhysRevD.87.103011} {\bibfield
  {journal} {\bibinfo  {journal} {Phys. Rev. D}\ }\textbf {\bibinfo {volume}
  {87}},\ \bibinfo {pages} {103011} (\bibinfo {year} {2013})}\BibitemShut
  {NoStop}%
\bibitem [{\citenamefont {Krippendorf}\ \emph {et~al.}(2018)\citenamefont
  {Krippendorf}, \citenamefont {Muia},\ and\ \citenamefont
  {Quevedo}}]{Krippendorf2018}%
  \BibitemOpen
  \bibfield  {author} {\bibinfo {author} {\bibfnamefont {S.}~\bibnamefont
  {Krippendorf}}, \bibinfo {author} {\bibfnamefont {F.}~\bibnamefont {Muia}}, \
  and\ \bibinfo {author} {\bibfnamefont {F.}~\bibnamefont {Quevedo}},\ }\href
  {\doibase 10.1007/JHEP08(2018)070} {\bibfield  {journal} {\bibinfo  {journal}
  {Journal of High Energy Physics}\ }\textbf {\bibinfo {volume} {2018}},\
  \bibinfo {pages} {70} (\bibinfo {year} {2018})},\ \Eprint
  {http://arxiv.org/abs/1806.04690} {arXiv:1806.04690} \BibitemShut {NoStop}%
\bibitem [{\citenamefont {Kibble}(1980)}]{Kibble1980}%
  \BibitemOpen
  \bibfield  {author} {\bibinfo {author} {\bibfnamefont {T.}~\bibnamefont
  {Kibble}},\ }\href {\doibase 10.1016/0370-1573(80)90091-5} {\bibfield
  {journal} {\bibinfo  {journal} {Physics Reports}\ }\textbf {\bibinfo {volume}
  {67}},\ \bibinfo {pages} {183} (\bibinfo {year} {1980})}\BibitemShut
  {NoStop}%
\bibitem [{\citenamefont {Vilenkin}(1985)}]{Vilenkin1985}%
  \BibitemOpen
  \bibfield  {author} {\bibinfo {author} {\bibfnamefont {A.}~\bibnamefont
  {Vilenkin}},\ }\href {\doibase 10.1016/0370-1573(85)90033-X} {\bibfield
  {journal} {\bibinfo  {journal} {Phys. Rep.}\ }\textbf {\bibinfo {volume}
  {121}},\ \bibinfo {pages} {263} (\bibinfo {year} {1985})}\BibitemShut
  {NoStop}%
\bibitem [{\citenamefont {Vilenkin}\ and\ \citenamefont
  {Shaposhnikov}(1994)}]{Vilenkin1994}%
  \BibitemOpen
  \bibfield  {author} {\bibinfo {author} {\bibfnamefont {A.}~\bibnamefont
  {Vilenkin}}\ and\ \bibinfo {author} {\bibfnamefont {M.}~\bibnamefont
  {Shaposhnikov}},\ }\href@noop {} {\emph {\bibinfo {title} {{Cosmic Strings
  and Other Topological Defects}}}}\ (\bibinfo  {publisher} {Cambridge
  University Press},\ \bibinfo {address} {Cambridge},\ \bibinfo {year}
  {1994})\BibitemShut {NoStop}%
\bibitem [{\citenamefont {Ge}\ \emph {et~al.}(2019)\citenamefont {Ge},
  \citenamefont {Lawson},\ and\ \citenamefont {Zhitnitsky}}]{Zhitnitsky2019}%
  \BibitemOpen
  \bibfield  {author} {\bibinfo {author} {\bibfnamefont {S.}~\bibnamefont
  {Ge}}, \bibinfo {author} {\bibfnamefont {K.}~\bibnamefont {Lawson}}, \ and\
  \bibinfo {author} {\bibfnamefont {A.}~\bibnamefont {Zhitnitsky}},\ }\href
  {\doibase 10.1103/PhysRevD.99.116017} {\bibfield  {journal} {\bibinfo
  {journal} {Phys. Rev. D}\ }\textbf {\bibinfo {volume} {99}},\ \bibinfo
  {pages} {116017} (\bibinfo {year} {2019})},\ \Eprint
  {http://arxiv.org/abs/1903.05090} {arXiv:1903.05090} \BibitemShut {NoStop}%
\bibitem [{\citenamefont {Budker}\ \emph
  {et~al.}(2020{\natexlab{a}})\citenamefont {Budker}, \citenamefont {Flambaum},
  \citenamefont {Liang},\ and\ \citenamefont {Zhitnitsky}}]{Budker2019}%
  \BibitemOpen
  \bibfield  {author} {\bibinfo {author} {\bibfnamefont {D.}~\bibnamefont
  {Budker}}, \bibinfo {author} {\bibfnamefont {V.~V.}\ \bibnamefont
  {Flambaum}}, \bibinfo {author} {\bibfnamefont {X.}~\bibnamefont {Liang}}, \
  and\ \bibinfo {author} {\bibfnamefont {A.}~\bibnamefont {Zhitnitsky}},\
  }\href {\doibase 10.1103/PhysRevD.101.043012} {\bibfield  {journal} {\bibinfo
   {journal} {Phys. Rev. D}\ }\textbf {\bibinfo {volume} {101}},\ \bibinfo
  {pages} {043012} (\bibinfo {year} {2020}{\natexlab{a}})},\ \Eprint
  {http://arxiv.org/abs/1909.09475} {arXiv:1909.09475} \BibitemShut {NoStop}%
\bibitem [{\citenamefont {Budker}\ \emph
  {et~al.}(2020{\natexlab{b}})\citenamefont {Budker}, \citenamefont
  {Flambaum},\ and\ \citenamefont {Zhitnitsky}}]{Budker2020a}%
  \BibitemOpen
  \bibfield  {author} {\bibinfo {author} {\bibfnamefont {D.}~\bibnamefont
  {Budker}}, \bibinfo {author} {\bibfnamefont {V.~V.}\ \bibnamefont
  {Flambaum}}, \ and\ \bibinfo {author} {\bibfnamefont {A.}~\bibnamefont
  {Zhitnitsky}},\ }\href {http://arxiv.org/abs/2003.07363} {\  (\bibinfo {year}
  {2020}{\natexlab{b}})},\ \Eprint {http://arxiv.org/abs/2003.07363}
  {arXiv:2003.07363} \BibitemShut {NoStop}%
\bibitem [{\citenamefont {Grabowska}\ \emph {et~al.}(2018)\citenamefont
  {Grabowska}, \citenamefont {Melia},\ and\ \citenamefont
  {Rajendran}}]{Grabowska2018}%
  \BibitemOpen
  \bibfield  {author} {\bibinfo {author} {\bibfnamefont {D.~M.}\ \bibnamefont
  {Grabowska}}, \bibinfo {author} {\bibfnamefont {T.}~\bibnamefont {Melia}}, \
  and\ \bibinfo {author} {\bibfnamefont {S.}~\bibnamefont {Rajendran}},\ }\href
  {\doibase 10.1103/PhysRevD.98.115020} {\bibfield  {journal} {\bibinfo
  {journal} {Physical Review D}\ }\textbf {\bibinfo {volume} {98}},\ \bibinfo
  {pages} {115020} (\bibinfo {year} {2018})},\ \Eprint
  {http://arxiv.org/abs/1807.03788} {arXiv:1807.03788} \BibitemShut {NoStop}%
\bibitem [{\citenamefont {Brax}\ \emph {et~al.}(2020)\citenamefont {Brax},
  \citenamefont {Valageas},\ and\ \citenamefont {Cembranos}}]{Brax2020}%
  \BibitemOpen
  \bibfield  {author} {\bibinfo {author} {\bibfnamefont {P.}~\bibnamefont
  {Brax}}, \bibinfo {author} {\bibfnamefont {P.}~\bibnamefont {Valageas}}, \
  and\ \bibinfo {author} {\bibfnamefont {J.~A.}\ \bibnamefont {Cembranos}},\
  }\href {\doibase 10.1103/PhysRevD.102.083012} {\bibfield  {journal} {\bibinfo
   {journal} {Phys. Rev. D}\ }\textbf {\bibinfo {volume} {102}},\ \bibinfo
  {pages} {22} (\bibinfo {year} {2020})},\ \Eprint
  {http://arxiv.org/abs/arXiv:2007.04638v3} {arXiv:arXiv:2007.04638v3}
  \BibitemShut {NoStop}%
\bibitem [{\citenamefont {Niikura}\ \emph {et~al.}(2019)\citenamefont
  {Niikura}, \citenamefont {Takada}, \citenamefont {Yasuda}, \citenamefont
  {Lupton}, \citenamefont {Sumi}, \citenamefont {More}, \citenamefont {Kurita},
  \citenamefont {Sugiyama}, \citenamefont {More}, \citenamefont {Oguri},\ and\
  \citenamefont {Chiba}}]{Niikura2019}%
  \BibitemOpen
  \bibfield  {author} {\bibinfo {author} {\bibfnamefont {H.}~\bibnamefont
  {Niikura}}, \bibinfo {author} {\bibfnamefont {M.}~\bibnamefont {Takada}},
  \bibinfo {author} {\bibfnamefont {N.}~\bibnamefont {Yasuda}}, \bibinfo
  {author} {\bibfnamefont {R.~H.}\ \bibnamefont {Lupton}}, \bibinfo {author}
  {\bibfnamefont {T.}~\bibnamefont {Sumi}}, \bibinfo {author} {\bibfnamefont
  {S.}~\bibnamefont {More}}, \bibinfo {author} {\bibfnamefont {T.}~\bibnamefont
  {Kurita}}, \bibinfo {author} {\bibfnamefont {S.}~\bibnamefont {Sugiyama}},
  \bibinfo {author} {\bibfnamefont {A.}~\bibnamefont {More}}, \bibinfo {author}
  {\bibfnamefont {M.}~\bibnamefont {Oguri}}, \ and\ \bibinfo {author}
  {\bibfnamefont {M.}~\bibnamefont {Chiba}},\ }\href {\doibase
  10.1038/s41550-019-0723-1} {\bibfield  {journal} {\bibinfo  {journal} {Nature
  Astronomy}\ }\textbf {\bibinfo {volume} {3}},\ \bibinfo {pages} {524}
  (\bibinfo {year} {2019})},\ \Eprint {http://arxiv.org/abs/1701.02151}
  {arXiv:1701.02151} \BibitemShut {NoStop}%
\bibitem [{\citenamefont {Nesti}\ and\ \citenamefont
  {Salucci}(2013)}]{NesSal13}%
  \BibitemOpen
  \bibfield  {author} {\bibinfo {author} {\bibfnamefont {F.}~\bibnamefont
  {Nesti}}\ and\ \bibinfo {author} {\bibfnamefont {P.}~\bibnamefont
  {Salucci}},\ }\href {http://arxiv.org/abs/1304.5127
  http://stacks.iop.org/1475-7516/2013/i=07/a=016} {\bibfield  {journal}
  {\bibinfo  {journal} {Journal of Cosmology and Astroparticle Physics}\
  }\textbf {\bibinfo {volume} {2013}},\ \bibinfo {pages} {16} (\bibinfo {year}
  {2013})},\ \Eprint {http://arxiv.org/abs/arXiv:1304.5127v2}
  {arXiv:arXiv:1304.5127v2} \BibitemShut {NoStop}%
\bibitem [{\citenamefont {Freese}\ \emph {et~al.}(2013)\citenamefont {Freese},
  \citenamefont {Lisanti},\ and\ \citenamefont {Savage}}]{Freese2013}%
  \BibitemOpen
  \bibfield  {author} {\bibinfo {author} {\bibfnamefont {K.}~\bibnamefont
  {Freese}}, \bibinfo {author} {\bibfnamefont {M.}~\bibnamefont {Lisanti}}, \
  and\ \bibinfo {author} {\bibfnamefont {C.}~\bibnamefont {Savage}},\ }\href
  {\doibase 10.1103/RevModPhys.85.1561} {\bibfield  {journal} {\bibinfo
  {journal} {Reviews of Modern Physics}\ }\textbf {\bibinfo {volume} {85}},\
  \bibinfo {pages} {1561} (\bibinfo {year} {2013})}\BibitemShut {NoStop}%
\bibitem [{\citenamefont {Pospelov}\ \emph {et~al.}(2013)\citenamefont
  {Pospelov}, \citenamefont {Pustelny}, \citenamefont {Ledbetter},
  \citenamefont {Kimball}, \citenamefont {Gawlik},\ and\ \citenamefont
  {Budker}}]{Pospelov2013}%
  \BibitemOpen
  \bibfield  {author} {\bibinfo {author} {\bibfnamefont {M.}~\bibnamefont
  {Pospelov}}, \bibinfo {author} {\bibfnamefont {S.}~\bibnamefont {Pustelny}},
  \bibinfo {author} {\bibfnamefont {M.~P.}\ \bibnamefont {Ledbetter}}, \bibinfo
  {author} {\bibfnamefont {D.~F.~J.}\ \bibnamefont {Kimball}}, \bibinfo
  {author} {\bibfnamefont {W.}~\bibnamefont {Gawlik}}, \ and\ \bibinfo {author}
  {\bibfnamefont {D.}~\bibnamefont {Budker}},\ }\href {\doibase
  10.1103/PhysRevLett.110.021803} {\bibfield  {journal} {\bibinfo  {journal}
  {Phys. Rev. Lett.}\ }\textbf {\bibinfo {volume} {110}},\ \bibinfo {pages}
  {021803} (\bibinfo {year} {2013})}\BibitemShut {NoStop}%
\bibitem [{\citenamefont {Roberts}\ \emph {et~al.}(2018)\citenamefont
  {Roberts}, \citenamefont {Blewitt}, \citenamefont {Dailey},\ and\
  \citenamefont {Derevianko}}]{GPSDM2018}%
  \BibitemOpen
  \bibfield  {author} {\bibinfo {author} {\bibfnamefont {B.~M.}\ \bibnamefont
  {Roberts}}, \bibinfo {author} {\bibfnamefont {G.}~\bibnamefont {Blewitt}},
  \bibinfo {author} {\bibfnamefont {C.}~\bibnamefont {Dailey}}, \ and\ \bibinfo
  {author} {\bibfnamefont {A.}~\bibnamefont {Derevianko}},\ }\href {\doibase
  10.1103/PhysRevD.97.083009} {\bibfield  {journal} {\bibinfo  {journal} {Phys.
  Rev. D}\ }\textbf {\bibinfo {volume} {97}},\ \bibinfo {pages} {083009}
  (\bibinfo {year} {2018})},\ \Eprint {http://arxiv.org/abs/1803.10264}
  {arXiv:1803.10264} \BibitemShut {NoStop}%
\bibitem [{\citenamefont {Panelli}\ \emph {et~al.}(2020)\citenamefont
  {Panelli}, \citenamefont {Roberts},\ and\ \citenamefont
  {Derevianko}}]{Panelli2019}%
  \BibitemOpen
  \bibfield  {author} {\bibinfo {author} {\bibfnamefont {G.}~\bibnamefont
  {Panelli}}, \bibinfo {author} {\bibfnamefont {B.~M.}\ \bibnamefont
  {Roberts}}, \ and\ \bibinfo {author} {\bibfnamefont {A.}~\bibnamefont
  {Derevianko}},\ }\href {\doibase 10.1140/epjqt/s40507-020-00081-9} {\bibfield
   {journal} {\bibinfo  {journal} {EPJ Quantum Technol.}\ }\textbf {\bibinfo
  {volume} {7}},\ \bibinfo {pages} {5} (\bibinfo {year} {2020})},\ \Eprint
  {http://arxiv.org/abs/1908.03320} {arXiv:1908.03320} \BibitemShut {NoStop}%
\bibitem [{\citenamefont {Masia-Roig}\ \emph {et~al.}(2020)\citenamefont
  {Masia-Roig}, \citenamefont {Smiga}, \citenamefont {Budker}, \citenamefont
  {Dumont}, \citenamefont {Grujic}, \citenamefont {Kim}, \citenamefont
  {{Jackson Kimball}}, \citenamefont {Lebedev}, \citenamefont {Monroy},
  \citenamefont {Pustelny}, \citenamefont {Scholtes}, \citenamefont {Segura},
  \citenamefont {Semertzidis}, \citenamefont {Shin}, \citenamefont {Stalnaker},
  \citenamefont {Sulai}, \citenamefont {Weis},\ and\ \citenamefont
  {Wickenbrock}}]{MasiaRoig2019}%
  \BibitemOpen
  \bibfield  {author} {\bibinfo {author} {\bibfnamefont {H.}~\bibnamefont
  {Masia-Roig}}, \bibinfo {author} {\bibfnamefont {J.~A.}\ \bibnamefont
  {Smiga}}, \bibinfo {author} {\bibfnamefont {D.}~\bibnamefont {Budker}},
  \bibinfo {author} {\bibfnamefont {V.}~\bibnamefont {Dumont}}, \bibinfo
  {author} {\bibfnamefont {Z.}~\bibnamefont {Grujic}}, \bibinfo {author}
  {\bibfnamefont {D.}~\bibnamefont {Kim}}, \bibinfo {author} {\bibfnamefont
  {D.~F.}\ \bibnamefont {{Jackson Kimball}}}, \bibinfo {author} {\bibfnamefont
  {V.}~\bibnamefont {Lebedev}}, \bibinfo {author} {\bibfnamefont
  {M.}~\bibnamefont {Monroy}}, \bibinfo {author} {\bibfnamefont
  {S.}~\bibnamefont {Pustelny}}, \bibinfo {author} {\bibfnamefont
  {T.}~\bibnamefont {Scholtes}}, \bibinfo {author} {\bibfnamefont {P.~C.}\
  \bibnamefont {Segura}}, \bibinfo {author} {\bibfnamefont {Y.~K.}\
  \bibnamefont {Semertzidis}}, \bibinfo {author} {\bibfnamefont {Y.~C.}\
  \bibnamefont {Shin}}, \bibinfo {author} {\bibfnamefont {J.~E.}\ \bibnamefont
  {Stalnaker}}, \bibinfo {author} {\bibfnamefont {I.}~\bibnamefont {Sulai}},
  \bibinfo {author} {\bibfnamefont {A.}~\bibnamefont {Weis}}, \ and\ \bibinfo
  {author} {\bibfnamefont {A.}~\bibnamefont {Wickenbrock}},\ }\href {\doibase
  10.1016/j.dark.2020.100494} {\bibfield  {journal} {\bibinfo  {journal} {Phys.
  Dark Universe}\ }\textbf {\bibinfo {volume} {28}},\ \bibinfo {pages} {100494}
  (\bibinfo {year} {2020})},\ \Eprint {http://arxiv.org/abs/1912.08727}
  {arXiv:1912.08727} \BibitemShut {NoStop}%
\bibitem [{\citenamefont {Jaeckel}\ \emph {et~al.}(2020)\citenamefont
  {Jaeckel}, \citenamefont {Schenk},\ and\ \citenamefont
  {Spannowsky}}]{Jaeckel2020}%
  \BibitemOpen
  \bibfield  {author} {\bibinfo {author} {\bibfnamefont {J.}~\bibnamefont
  {Jaeckel}}, \bibinfo {author} {\bibfnamefont {S.}~\bibnamefont {Schenk}}, \
  and\ \bibinfo {author} {\bibfnamefont {M.}~\bibnamefont {Spannowsky}},\
  }\href {http://arxiv.org/abs/2004.13724} {\  (\bibinfo {year} {2020})},\
  \Eprint {http://arxiv.org/abs/2004.13724} {arXiv:2004.13724} \BibitemShut
  {NoStop}%
\bibitem [{\citenamefont {Dailey}\ \emph {et~al.}(2020)\citenamefont {Dailey},
  \citenamefont {Bradley}, \citenamefont {{Jackson Kimball}}, \citenamefont
  {Sulai}, \citenamefont {Pustelny}, \citenamefont {Wickenbrock},\ and\
  \citenamefont {Derevianko}}]{Dailey2020}%
  \BibitemOpen
  \bibfield  {author} {\bibinfo {author} {\bibfnamefont {C.}~\bibnamefont
  {Dailey}}, \bibinfo {author} {\bibfnamefont {C.}~\bibnamefont {Bradley}},
  \bibinfo {author} {\bibfnamefont {D.~F.}\ \bibnamefont {{Jackson Kimball}}},
  \bibinfo {author} {\bibfnamefont {I.~A.}\ \bibnamefont {Sulai}}, \bibinfo
  {author} {\bibfnamefont {S.}~\bibnamefont {Pustelny}}, \bibinfo {author}
  {\bibfnamefont {A.}~\bibnamefont {Wickenbrock}}, \ and\ \bibinfo {author}
  {\bibfnamefont {A.}~\bibnamefont {Derevianko}},\ }\href {\doibase
  10.1038/s41550-020-01242-7} {\bibfield  {journal} {\bibinfo  {journal} {Nat.
  Astron.}\ } (\bibinfo {year} {2020}),\ 10.1038/s41550-020-01242-7},\ \Eprint
  {http://arxiv.org/abs/2002.04352} {arXiv:2002.04352} \BibitemShut {NoStop}%
\bibitem [{\citenamefont {{The LIGO Scientific Collaboration and Virgo
  Collaboration}}(2016)}]{LIGO2016}%
  \BibitemOpen
  \bibfield  {author} {\bibinfo {author} {\bibnamefont {{The LIGO Scientific
  Collaboration and Virgo Collaboration}}},\ }\href {\doibase
  10.1103/PhysRevLett.116.061102} {\bibfield  {journal} {\bibinfo  {journal}
  {Phys. Rev. Lett.}\ }\textbf {\bibinfo {volume} {116}},\ \bibinfo {pages}
  {061102} (\bibinfo {year} {2016})}\BibitemShut {NoStop}%
\bibitem [{\citenamefont {Arvanitaki}\ \emph {et~al.}(2015)\citenamefont
  {Arvanitaki}, \citenamefont {Huang},\ and\ \citenamefont {{Van
  Tilburg}}}]{Arvanitaki2014}%
  \BibitemOpen
  \bibfield  {author} {\bibinfo {author} {\bibfnamefont {A.}~\bibnamefont
  {Arvanitaki}}, \bibinfo {author} {\bibfnamefont {J.}~\bibnamefont {Huang}}, \
  and\ \bibinfo {author} {\bibfnamefont {K.}~\bibnamefont {{Van Tilburg}}},\
  }\href {\doibase 10.1103/PhysRevD.91.015015} {\bibfield  {journal} {\bibinfo
  {journal} {Phys. Rev. D}\ }\textbf {\bibinfo {volume} {91}},\ \bibinfo
  {pages} {015015} (\bibinfo {year} {2015})}\BibitemShut {NoStop}%
\bibitem [{\citenamefont {{Van Tilburg}}\ \emph {et~al.}(2015)\citenamefont
  {{Van Tilburg}}, \citenamefont {Leefer}, \citenamefont {Bougas},\ and\
  \citenamefont {Budker}}]{Tilburg2015}%
  \BibitemOpen
  \bibfield  {author} {\bibinfo {author} {\bibfnamefont {K.}~\bibnamefont {{Van
  Tilburg}}}, \bibinfo {author} {\bibfnamefont {N.}~\bibnamefont {Leefer}},
  \bibinfo {author} {\bibfnamefont {L.}~\bibnamefont {Bougas}}, \ and\ \bibinfo
  {author} {\bibfnamefont {D.}~\bibnamefont {Budker}},\ }\href {\doibase
  10.1103/PhysRevLett.115.011802} {\bibfield  {journal} {\bibinfo  {journal}
  {Phys. Rev. Lett.}\ }\textbf {\bibinfo {volume} {115}},\ \bibinfo {pages}
  {011802} (\bibinfo {year} {2015})}\BibitemShut {NoStop}%
\bibitem [{\citenamefont {Hees}\ \emph {et~al.}(2016)\citenamefont {Hees},
  \citenamefont {Gu{\'{e}}na}, \citenamefont {Abgrall}, \citenamefont {Bize},\
  and\ \citenamefont {Wolf}}]{Hees2016}%
  \BibitemOpen
  \bibfield  {author} {\bibinfo {author} {\bibfnamefont {A.}~\bibnamefont
  {Hees}}, \bibinfo {author} {\bibfnamefont {J.}~\bibnamefont {Gu{\'{e}}na}},
  \bibinfo {author} {\bibfnamefont {M.}~\bibnamefont {Abgrall}}, \bibinfo
  {author} {\bibfnamefont {S.}~\bibnamefont {Bize}}, \ and\ \bibinfo {author}
  {\bibfnamefont {P.}~\bibnamefont {Wolf}},\ }\href {\doibase
  10.1103/PhysRevLett.117.061301} {\bibfield  {journal} {\bibinfo  {journal}
  {Phys. Rev. Lett.}\ }\textbf {\bibinfo {volume} {117}},\ \bibinfo {pages}
  {061301} (\bibinfo {year} {2016})},\ \Eprint
  {http://arxiv.org/abs/1604.08514} {arXiv:1604.08514 [gr-qc]} \BibitemShut
  {NoStop}%
\bibitem [{\citenamefont {Wcis{\l}o}\ \emph {et~al.}(2016)\citenamefont
  {Wcis{\l}o}, \citenamefont {Morzy{\'{n}}ski}, \citenamefont {Bober},
  \citenamefont {Cygan}, \citenamefont {Lisak}, \citenamefont {Ciury{\l}o},\
  and\ \citenamefont {Zawada}}]{Wcislo2016}%
  \BibitemOpen
  \bibfield  {author} {\bibinfo {author} {\bibfnamefont {P.}~\bibnamefont
  {Wcis{\l}o}}, \bibinfo {author} {\bibfnamefont {P.}~\bibnamefont
  {Morzy{\'{n}}ski}}, \bibinfo {author} {\bibfnamefont {M.}~\bibnamefont
  {Bober}}, \bibinfo {author} {\bibfnamefont {A.}~\bibnamefont {Cygan}},
  \bibinfo {author} {\bibfnamefont {D.}~\bibnamefont {Lisak}}, \bibinfo
  {author} {\bibfnamefont {R.}~\bibnamefont {Ciury{\l}o}}, \ and\ \bibinfo
  {author} {\bibfnamefont {M.}~\bibnamefont {Zawada}},\ }\href {\doibase
  10.1038/s41550-016-0009} {\bibfield  {journal} {\bibinfo  {journal} {Nature
  Astronomy}\ }\textbf {\bibinfo {volume} {1}},\ \bibinfo {pages} {0009}
  (\bibinfo {year} {2016})}\BibitemShut {NoStop}%
\bibitem [{\citenamefont {Roberts}\ \emph {et~al.}(2017)\citenamefont
  {Roberts}, \citenamefont {Blewitt}, \citenamefont {Dailey}, \citenamefont
  {Murphy}, \citenamefont {Pospelov}, \citenamefont {Rollings}, \citenamefont
  {Sherman}, \citenamefont {Williams},\ and\ \citenamefont
  {Derevianko}}]{GPSDM2017}%
  \BibitemOpen
  \bibfield  {author} {\bibinfo {author} {\bibfnamefont {B.~M.}\ \bibnamefont
  {Roberts}}, \bibinfo {author} {\bibfnamefont {G.}~\bibnamefont {Blewitt}},
  \bibinfo {author} {\bibfnamefont {C.}~\bibnamefont {Dailey}}, \bibinfo
  {author} {\bibfnamefont {M.}~\bibnamefont {Murphy}}, \bibinfo {author}
  {\bibfnamefont {M.}~\bibnamefont {Pospelov}}, \bibinfo {author}
  {\bibfnamefont {A.}~\bibnamefont {Rollings}}, \bibinfo {author}
  {\bibfnamefont {J.}~\bibnamefont {Sherman}}, \bibinfo {author} {\bibfnamefont
  {W.}~\bibnamefont {Williams}}, \ and\ \bibinfo {author} {\bibfnamefont
  {A.}~\bibnamefont {Derevianko}},\ }\href {\doibase
  10.1038/s41467-017-01440-4} {\bibfield  {journal} {\bibinfo  {journal} {Nat.
  Commun.}\ }\textbf {\bibinfo {volume} {8}},\ \bibinfo {pages} {1195}
  (\bibinfo {year} {2017})},\ \Eprint {http://arxiv.org/abs/1704.06844}
  {arXiv:1704.06844} \BibitemShut {NoStop}%
\bibitem [{\citenamefont {Kalaydzhyan}\ and\ \citenamefont
  {Yu}(2017)}]{Kalaydzhyan2017}%
  \BibitemOpen
  \bibfield  {author} {\bibinfo {author} {\bibfnamefont {T.}~\bibnamefont
  {Kalaydzhyan}}\ and\ \bibinfo {author} {\bibfnamefont {N.}~\bibnamefont
  {Yu}},\ }\href {http://arxiv.org/abs/1705.05833} {\  (\bibinfo {year}
  {2017})},\ \Eprint {http://arxiv.org/abs/1705.05833} {arXiv:1705.05833}
  \BibitemShut {NoStop}%
\bibitem [{\citenamefont {Wcislo}\ \emph {et~al.}(2018)\citenamefont {Wcislo},
  \citenamefont {Ablewski}, \citenamefont {Beloy}, \citenamefont {Bilicki},
  \citenamefont {Bober}, \citenamefont {Brown}, \citenamefont {Fasano},
  \citenamefont {Ciurylo}, \citenamefont {Hachisu}, \citenamefont {Ido},
  \citenamefont {Lodewyck}, \citenamefont {Ludlow}, \citenamefont {McGrew},
  \citenamefont {Morzy{\'{n}}ski}, \citenamefont {Nicolodi}, \citenamefont
  {Schioppo}, \citenamefont {Sekido}, \citenamefont {{Le Targat}},
  \citenamefont {Wolf}, \citenamefont {Zhang}, \citenamefont {Zjawin},
  \citenamefont {Zawada}, \citenamefont {Wcis{\l}o}, \citenamefont {Ablewski},
  \citenamefont {Beloy}, \citenamefont {Bilicki}, \citenamefont {Bober},
  \citenamefont {Brown}, \citenamefont {Fasano}, \citenamefont {Ciury{\l}o},
  \citenamefont {Hachisu}, \citenamefont {Ido}, \citenamefont {Lodewyck},
  \citenamefont {Ludlow}, \citenamefont {McGrew}, \citenamefont
  {Morzy{\'{n}}ski}, \citenamefont {Nicolodi}, \citenamefont {Schioppo},
  \citenamefont {Sekido}, \citenamefont {{Le Targat}}, \citenamefont {Wolf},
  \citenamefont {Zhang}, \citenamefont {Zjawin},\ and\ \citenamefont
  {Zawada}}]{Wcislo-clock-network-2018}%
  \BibitemOpen
  \bibfield  {author} {\bibinfo {author} {\bibfnamefont {P.}~\bibnamefont
  {Wcislo}}, \bibinfo {author} {\bibfnamefont {P.}~\bibnamefont {Ablewski}},
  \bibinfo {author} {\bibfnamefont {K.}~\bibnamefont {Beloy}}, \bibinfo
  {author} {\bibfnamefont {S.}~\bibnamefont {Bilicki}}, \bibinfo {author}
  {\bibfnamefont {M.}~\bibnamefont {Bober}}, \bibinfo {author} {\bibfnamefont
  {R.}~\bibnamefont {Brown}}, \bibinfo {author} {\bibfnamefont
  {R.}~\bibnamefont {Fasano}}, \bibinfo {author} {\bibfnamefont
  {R.}~\bibnamefont {Ciurylo}}, \bibinfo {author} {\bibfnamefont
  {H.}~\bibnamefont {Hachisu}}, \bibinfo {author} {\bibfnamefont
  {T.}~\bibnamefont {Ido}}, \bibinfo {author} {\bibfnamefont {J.}~\bibnamefont
  {Lodewyck}}, \bibinfo {author} {\bibfnamefont {A.}~\bibnamefont {Ludlow}},
  \bibinfo {author} {\bibfnamefont {W.}~\bibnamefont {McGrew}}, \bibinfo
  {author} {\bibfnamefont {P.}~\bibnamefont {Morzy{\'{n}}ski}}, \bibinfo
  {author} {\bibfnamefont {D.}~\bibnamefont {Nicolodi}}, \bibinfo {author}
  {\bibfnamefont {M.}~\bibnamefont {Schioppo}}, \bibinfo {author}
  {\bibfnamefont {M.}~\bibnamefont {Sekido}}, \bibinfo {author} {\bibfnamefont
  {R.}~\bibnamefont {{Le Targat}}}, \bibinfo {author} {\bibfnamefont
  {P.}~\bibnamefont {Wolf}}, \bibinfo {author} {\bibfnamefont {X.}~\bibnamefont
  {Zhang}}, \bibinfo {author} {\bibfnamefont {B.}~\bibnamefont {Zjawin}},
  \bibinfo {author} {\bibfnamefont {M.}~\bibnamefont {Zawada}}, \bibinfo
  {author} {\bibfnamefont {P.}~\bibnamefont {Wcis{\l}o}}, \bibinfo {author}
  {\bibfnamefont {P.}~\bibnamefont {Ablewski}}, \bibinfo {author}
  {\bibfnamefont {K.}~\bibnamefont {Beloy}}, \bibinfo {author} {\bibfnamefont
  {S.}~\bibnamefont {Bilicki}}, \bibinfo {author} {\bibfnamefont
  {M.}~\bibnamefont {Bober}}, \bibinfo {author} {\bibfnamefont
  {R.}~\bibnamefont {Brown}}, \bibinfo {author} {\bibfnamefont
  {R.}~\bibnamefont {Fasano}}, \bibinfo {author} {\bibfnamefont
  {R.}~\bibnamefont {Ciury{\l}o}}, \bibinfo {author} {\bibfnamefont
  {H.}~\bibnamefont {Hachisu}}, \bibinfo {author} {\bibfnamefont
  {T.}~\bibnamefont {Ido}}, \bibinfo {author} {\bibfnamefont {J.}~\bibnamefont
  {Lodewyck}}, \bibinfo {author} {\bibfnamefont {A.}~\bibnamefont {Ludlow}},
  \bibinfo {author} {\bibfnamefont {W.}~\bibnamefont {McGrew}}, \bibinfo
  {author} {\bibfnamefont {P.}~\bibnamefont {Morzy{\'{n}}ski}}, \bibinfo
  {author} {\bibfnamefont {D.}~\bibnamefont {Nicolodi}}, \bibinfo {author}
  {\bibfnamefont {M.}~\bibnamefont {Schioppo}}, \bibinfo {author}
  {\bibfnamefont {M.}~\bibnamefont {Sekido}}, \bibinfo {author} {\bibfnamefont
  {R.}~\bibnamefont {{Le Targat}}}, \bibinfo {author} {\bibfnamefont
  {P.}~\bibnamefont {Wolf}}, \bibinfo {author} {\bibfnamefont {X.}~\bibnamefont
  {Zhang}}, \bibinfo {author} {\bibfnamefont {B.}~\bibnamefont {Zjawin}}, \
  and\ \bibinfo {author} {\bibfnamefont {M.}~\bibnamefont {Zawada}},\ }\href
  {\doibase 10.1126/sciadv.aau4869} {\bibfield  {journal} {\bibinfo  {journal}
  {Science Advances}\ }\textbf {\bibinfo {volume} {4}},\ \bibinfo {pages}
  {eaau4869} (\bibinfo {year} {2018})}\BibitemShut {NoStop}%
\bibitem [{\citenamefont {Savalle}\ \emph {et~al.}(2021)\citenamefont
  {Savalle}, \citenamefont {Hees}, \citenamefont {Frank}, \citenamefont
  {Cantin}, \citenamefont {Pottie}, \citenamefont {Roberts}, \citenamefont
  {Cros}, \citenamefont {McAllister},\ and\ \citenamefont
  {Wolf}}]{Savalle2020}%
  \BibitemOpen
  \bibfield  {author} {\bibinfo {author} {\bibfnamefont {E.}~\bibnamefont
  {Savalle}}, \bibinfo {author} {\bibfnamefont {A.}~\bibnamefont {Hees}},
  \bibinfo {author} {\bibfnamefont {F.}~\bibnamefont {Frank}}, \bibinfo
  {author} {\bibfnamefont {E.}~\bibnamefont {Cantin}}, \bibinfo {author}
  {\bibfnamefont {P.-E.}\ \bibnamefont {Pottie}}, \bibinfo {author}
  {\bibfnamefont {B.~M.}\ \bibnamefont {Roberts}}, \bibinfo {author}
  {\bibfnamefont {L.}~\bibnamefont {Cros}}, \bibinfo {author} {\bibfnamefont
  {B.~T.}\ \bibnamefont {McAllister}}, \ and\ \bibinfo {author} {\bibfnamefont
  {P.}~\bibnamefont {Wolf}},\ }\href {http://arxiv.org/abs/2006.07055
  https://journals.aps.org/prl/accepted/27071Ye3Gae1b37452210073568beacb2f1667df8}
  {\bibfield  {journal} {\bibinfo  {journal} {Phys. Rev. Lett.}\ ,\ \bibinfo
  {pages} {accepted}} (\bibinfo {year} {2021})},\ \Eprint
  {http://arxiv.org/abs/2006.07055} {arXiv:2006.07055} \BibitemShut {NoStop}%
\bibitem [{\citenamefont {Roberts}\ \emph {et~al.}(2020)\citenamefont
  {Roberts}, \citenamefont {Delva}, \citenamefont {Al-Masoudi}, \citenamefont
  {Amy-Klein}, \citenamefont {B{\ae}rentsen}, \citenamefont {Baynham},
  \citenamefont {Benkler}, \citenamefont {Bilicki}, \citenamefont {Bize},
  \citenamefont {Bowden}, \citenamefont {Calvert}, \citenamefont {Cambier},
  \citenamefont {Cantin}, \citenamefont {Curtis}, \citenamefont
  {D{\"{o}}rscher}, \citenamefont {Favier}, \citenamefont {Frank},
  \citenamefont {Gill}, \citenamefont {Godun}, \citenamefont {Grosche},
  \citenamefont {Guo}, \citenamefont {Hees}, \citenamefont {Hill},
  \citenamefont {Hobson}, \citenamefont {Huntemann}, \citenamefont
  {Kronj{\"{a}}ger}, \citenamefont {Koke}, \citenamefont {Kuhl}, \citenamefont
  {Lange}, \citenamefont {Legero}, \citenamefont {Lipphardt}, \citenamefont
  {Lisdat}, \citenamefont {Lodewyck}, \citenamefont {Lopez}, \citenamefont
  {Margolis}, \citenamefont {{\'{A}}lvarez-Mart{\'{i}}nez}, \citenamefont
  {Meynadier}, \citenamefont {Ozimek}, \citenamefont {Peik}, \citenamefont
  {Pottie}, \citenamefont {Quintin}, \citenamefont {Sanner}, \citenamefont {{De
  Sarlo}}, \citenamefont {Schioppo}, \citenamefont {Schwarz}, \citenamefont
  {Silva}, \citenamefont {Sterr}, \citenamefont {Tamm}, \citenamefont {{Le
  Targat}}, \citenamefont {Tuckey}, \citenamefont {Vallet}, \citenamefont
  {Waterholter}, \citenamefont {Xu},\ and\ \citenamefont
  {Wolf}}]{RobertsTDs2020}%
  \BibitemOpen
  \bibfield  {author} {\bibinfo {author} {\bibfnamefont {B.~M.}\ \bibnamefont
  {Roberts}}, \bibinfo {author} {\bibfnamefont {P.}~\bibnamefont {Delva}},
  \bibinfo {author} {\bibfnamefont {A.}~\bibnamefont {Al-Masoudi}}, \bibinfo
  {author} {\bibfnamefont {A.}~\bibnamefont {Amy-Klein}}, \bibinfo {author}
  {\bibfnamefont {C.}~\bibnamefont {B{\ae}rentsen}}, \bibinfo {author}
  {\bibfnamefont {C.~F.~A.}\ \bibnamefont {Baynham}}, \bibinfo {author}
  {\bibfnamefont {E.}~\bibnamefont {Benkler}}, \bibinfo {author} {\bibfnamefont
  {S.}~\bibnamefont {Bilicki}}, \bibinfo {author} {\bibfnamefont
  {S.}~\bibnamefont {Bize}}, \bibinfo {author} {\bibfnamefont {W.}~\bibnamefont
  {Bowden}}, \bibinfo {author} {\bibfnamefont {J.}~\bibnamefont {Calvert}},
  \bibinfo {author} {\bibfnamefont {V.}~\bibnamefont {Cambier}}, \bibinfo
  {author} {\bibfnamefont {E.}~\bibnamefont {Cantin}}, \bibinfo {author}
  {\bibfnamefont {E.~A.}\ \bibnamefont {Curtis}}, \bibinfo {author}
  {\bibfnamefont {S.}~\bibnamefont {D{\"{o}}rscher}}, \bibinfo {author}
  {\bibfnamefont {M.}~\bibnamefont {Favier}}, \bibinfo {author} {\bibfnamefont
  {F.}~\bibnamefont {Frank}}, \bibinfo {author} {\bibfnamefont
  {P.}~\bibnamefont {Gill}}, \bibinfo {author} {\bibfnamefont {R.~M.}\
  \bibnamefont {Godun}}, \bibinfo {author} {\bibfnamefont {G.}~\bibnamefont
  {Grosche}}, \bibinfo {author} {\bibfnamefont {C.}~\bibnamefont {Guo}},
  \bibinfo {author} {\bibfnamefont {A.}~\bibnamefont {Hees}}, \bibinfo {author}
  {\bibfnamefont {I.~R.}\ \bibnamefont {Hill}}, \bibinfo {author}
  {\bibfnamefont {R.}~\bibnamefont {Hobson}}, \bibinfo {author} {\bibfnamefont
  {N.}~\bibnamefont {Huntemann}}, \bibinfo {author} {\bibfnamefont
  {J.}~\bibnamefont {Kronj{\"{a}}ger}}, \bibinfo {author} {\bibfnamefont
  {S.}~\bibnamefont {Koke}}, \bibinfo {author} {\bibfnamefont {A.}~\bibnamefont
  {Kuhl}}, \bibinfo {author} {\bibfnamefont {R.}~\bibnamefont {Lange}},
  \bibinfo {author} {\bibfnamefont {T.}~\bibnamefont {Legero}}, \bibinfo
  {author} {\bibfnamefont {B.}~\bibnamefont {Lipphardt}}, \bibinfo {author}
  {\bibfnamefont {C.}~\bibnamefont {Lisdat}}, \bibinfo {author} {\bibfnamefont
  {J.}~\bibnamefont {Lodewyck}}, \bibinfo {author} {\bibfnamefont
  {O.}~\bibnamefont {Lopez}}, \bibinfo {author} {\bibfnamefont {H.~S.}\
  \bibnamefont {Margolis}}, \bibinfo {author} {\bibfnamefont {H.}~\bibnamefont
  {{\'{A}}lvarez-Mart{\'{i}}nez}}, \bibinfo {author} {\bibfnamefont
  {F.}~\bibnamefont {Meynadier}}, \bibinfo {author} {\bibfnamefont
  {F.}~\bibnamefont {Ozimek}}, \bibinfo {author} {\bibfnamefont
  {E.}~\bibnamefont {Peik}}, \bibinfo {author} {\bibfnamefont {P.-E.}\
  \bibnamefont {Pottie}}, \bibinfo {author} {\bibfnamefont {N.}~\bibnamefont
  {Quintin}}, \bibinfo {author} {\bibfnamefont {C.}~\bibnamefont {Sanner}},
  \bibinfo {author} {\bibfnamefont {L.}~\bibnamefont {{De Sarlo}}}, \bibinfo
  {author} {\bibfnamefont {M.}~\bibnamefont {Schioppo}}, \bibinfo {author}
  {\bibfnamefont {R.}~\bibnamefont {Schwarz}}, \bibinfo {author} {\bibfnamefont
  {A.}~\bibnamefont {Silva}}, \bibinfo {author} {\bibfnamefont
  {U.}~\bibnamefont {Sterr}}, \bibinfo {author} {\bibfnamefont
  {C.}~\bibnamefont {Tamm}}, \bibinfo {author} {\bibfnamefont {R.}~\bibnamefont
  {{Le Targat}}}, \bibinfo {author} {\bibfnamefont {P.}~\bibnamefont {Tuckey}},
  \bibinfo {author} {\bibfnamefont {G.}~\bibnamefont {Vallet}}, \bibinfo
  {author} {\bibfnamefont {T.}~\bibnamefont {Waterholter}}, \bibinfo {author}
  {\bibfnamefont {D.}~\bibnamefont {Xu}}, \ and\ \bibinfo {author}
  {\bibfnamefont {P.}~\bibnamefont {Wolf}},\ }\href {\doibase
  10.1088/1367-2630/abaace} {\bibfield  {journal} {\bibinfo  {journal} {New J.
  Phys.}\ }\textbf {\bibinfo {volume} {22}},\ \bibinfo {pages} {093010}
  (\bibinfo {year} {2020})},\ \Eprint {http://arxiv.org/abs/1907.02661}
  {arXiv:1907.02661} \BibitemShut {NoStop}%
\bibitem [{\citenamefont {Stadnik}\ and\ \citenamefont
  {Flambaum}(2015)}]{StadnikLaser2015}%
  \BibitemOpen
  \bibfield  {author} {\bibinfo {author} {\bibfnamefont {Y.~V.}\ \bibnamefont
  {Stadnik}}\ and\ \bibinfo {author} {\bibfnamefont {V.~V.}\ \bibnamefont
  {Flambaum}},\ }\href {\doibase 10.1103/PhysRevLett.114.161301} {\bibfield
  {journal} {\bibinfo  {journal} {Phys. Rev. Lett.}\ }\textbf {\bibinfo
  {volume} {114}},\ \bibinfo {pages} {161301} (\bibinfo {year}
  {2015})}\BibitemShut {NoStop}%
\bibitem [{\citenamefont {Stadnik}\ and\ \citenamefont
  {Flambaum}(2016)}]{StadnikLasInf2015}%
  \BibitemOpen
  \bibfield  {author} {\bibinfo {author} {\bibfnamefont {Y.~V.}\ \bibnamefont
  {Stadnik}}\ and\ \bibinfo {author} {\bibfnamefont {V.~V.}\ \bibnamefont
  {Flambaum}},\ }\href {\doibase 10.1103/PhysRevA.93.063630} {\bibfield
  {journal} {\bibinfo  {journal} {Phys. Rev. A}\ }\textbf {\bibinfo {volume}
  {93}},\ \bibinfo {pages} {063630} (\bibinfo {year} {2016})}\BibitemShut
  {NoStop}%
\bibitem [{\citenamefont {Arvanitaki}\ \emph {et~al.}(2018)\citenamefont
  {Arvanitaki}, \citenamefont {Graham}, \citenamefont {Hogan}, \citenamefont
  {Rajendran},\ and\ \citenamefont {{Van Tilburg}}}]{Arvanitaki2018}%
  \BibitemOpen
  \bibfield  {author} {\bibinfo {author} {\bibfnamefont {A.}~\bibnamefont
  {Arvanitaki}}, \bibinfo {author} {\bibfnamefont {P.~W.}\ \bibnamefont
  {Graham}}, \bibinfo {author} {\bibfnamefont {J.~M.}\ \bibnamefont {Hogan}},
  \bibinfo {author} {\bibfnamefont {S.}~\bibnamefont {Rajendran}}, \ and\
  \bibinfo {author} {\bibfnamefont {K.}~\bibnamefont {{Van Tilburg}}},\ }\href
  {\doibase 10.1103/PhysRevD.97.075020} {\bibfield  {journal} {\bibinfo
  {journal} {Phys. Rev. D}\ }\textbf {\bibinfo {volume} {97}},\ \bibinfo
  {pages} {075020} (\bibinfo {year} {2018})},\ \Eprint
  {http://arxiv.org/abs/1606.04541} {arXiv:1606.04541} \BibitemShut {NoStop}%
\bibitem [{\citenamefont {Hu}\ \emph {et~al.}(2019)\citenamefont {Hu},
  \citenamefont {Lawson}, \citenamefont {Budker}, \citenamefont {Figueroa},
  \citenamefont {Kimball}, \citenamefont {Mills},\ and\ \citenamefont
  {Voigt}}]{Hu2019}%
  \BibitemOpen
  \bibfield  {author} {\bibinfo {author} {\bibfnamefont {W.}~\bibnamefont
  {Hu}}, \bibinfo {author} {\bibfnamefont {M.}~\bibnamefont {Lawson}}, \bibinfo
  {author} {\bibfnamefont {D.}~\bibnamefont {Budker}}, \bibinfo {author}
  {\bibfnamefont {N.~L.}\ \bibnamefont {Figueroa}}, \bibinfo {author}
  {\bibfnamefont {D.~F.~J.}\ \bibnamefont {Kimball}}, \bibinfo {author}
  {\bibfnamefont {A.~P.}\ \bibnamefont {Mills}}, \ and\ \bibinfo {author}
  {\bibfnamefont {C.}~\bibnamefont {Voigt}},\ }\href
  {http://arxiv.org/abs/1912.01900} {\  (\bibinfo {year} {2019})},\ \Eprint
  {http://arxiv.org/abs/1912.01900} {arXiv:1912.01900} \BibitemShut {NoStop}%
\bibitem [{\citenamefont {McNally}\ and\ \citenamefont
  {Zelevinsky}(2020)}]{McNally2019}%
  \BibitemOpen
  \bibfield  {author} {\bibinfo {author} {\bibfnamefont {R.~L.}\ \bibnamefont
  {McNally}}\ and\ \bibinfo {author} {\bibfnamefont {T.}~\bibnamefont
  {Zelevinsky}},\ }\href {\doibase 10.1140/epjd/e2020-100632-0} {\bibfield
  {journal} {\bibinfo  {journal} {Eur. Phys. J. D}\ }\textbf {\bibinfo {volume}
  {74}},\ \bibinfo {pages} {61} (\bibinfo {year} {2020})},\ \Eprint
  {http://arxiv.org/abs/1912.06703} {arXiv:1912.06703} \BibitemShut {NoStop}%
\bibitem [{\citenamefont {Savalle}\ \emph {et~al.}(2019)\citenamefont
  {Savalle}, \citenamefont {Roberts}, \citenamefont {Frank}, \citenamefont
  {Pottie}, \citenamefont {McAllister}, \citenamefont {Dailey}, \citenamefont
  {Derevianko},\ and\ \citenamefont {Wolf}}]{Savalle2019}%
  \BibitemOpen
  \bibfield  {author} {\bibinfo {author} {\bibfnamefont {E.}~\bibnamefont
  {Savalle}}, \bibinfo {author} {\bibfnamefont {B.~M.}\ \bibnamefont
  {Roberts}}, \bibinfo {author} {\bibfnamefont {F.}~\bibnamefont {Frank}},
  \bibinfo {author} {\bibfnamefont {P.-E.}\ \bibnamefont {Pottie}}, \bibinfo
  {author} {\bibfnamefont {B.~T.}\ \bibnamefont {McAllister}}, \bibinfo
  {author} {\bibfnamefont {C.}~\bibnamefont {Dailey}}, \bibinfo {author}
  {\bibfnamefont {A.}~\bibnamefont {Derevianko}}, \ and\ \bibinfo {author}
  {\bibfnamefont {P.}~\bibnamefont {Wolf}},\ }\href
  {https://arxiv.org/abs/1902.07192} {\  (\bibinfo {year} {2019})},\ \Eprint
  {http://arxiv.org/abs/1902.07192} {arXiv:1902.07192} \BibitemShut {NoStop}%
\bibitem [{\citenamefont {Baron}\ \emph {et~al.}(2014)\citenamefont {Baron},
  \citenamefont {Campbell}, \citenamefont {DeMille}, \citenamefont {Doyle},
  \citenamefont {Gabrielse}, \citenamefont {Gurevich}, \citenamefont {Hess},
  \citenamefont {Hutzler}, \citenamefont {Kirilov}, \citenamefont {Kozyryev},
  \citenamefont {O'Leary}, \citenamefont {Panda}, \citenamefont {Parsons},
  \citenamefont {Petrik}, \citenamefont {Spaun}, \citenamefont {Vutha},
  \citenamefont {West},\ and\ \citenamefont {{The ACME
  Collaboration}}}]{ACME2014}%
  \BibitemOpen
  \bibfield  {author} {\bibinfo {author} {\bibfnamefont {J.}~\bibnamefont
  {Baron}}, \bibinfo {author} {\bibfnamefont {W.~C.}\ \bibnamefont {Campbell}},
  \bibinfo {author} {\bibfnamefont {D.}~\bibnamefont {DeMille}}, \bibinfo
  {author} {\bibfnamefont {J.~M.}\ \bibnamefont {Doyle}}, \bibinfo {author}
  {\bibfnamefont {G.}~\bibnamefont {Gabrielse}}, \bibinfo {author}
  {\bibfnamefont {Y.~V.}\ \bibnamefont {Gurevich}}, \bibinfo {author}
  {\bibfnamefont {P.~W.}\ \bibnamefont {Hess}}, \bibinfo {author}
  {\bibfnamefont {N.~R.}\ \bibnamefont {Hutzler}}, \bibinfo {author}
  {\bibfnamefont {E.}~\bibnamefont {Kirilov}}, \bibinfo {author} {\bibfnamefont
  {I.}~\bibnamefont {Kozyryev}}, \bibinfo {author} {\bibfnamefont {B.~R.}\
  \bibnamefont {O'Leary}}, \bibinfo {author} {\bibfnamefont {C.~D.}\
  \bibnamefont {Panda}}, \bibinfo {author} {\bibfnamefont {M.~F.}\ \bibnamefont
  {Parsons}}, \bibinfo {author} {\bibfnamefont {E.~S.}\ \bibnamefont {Petrik}},
  \bibinfo {author} {\bibfnamefont {B.}~\bibnamefont {Spaun}}, \bibinfo
  {author} {\bibfnamefont {A.~C.}\ \bibnamefont {Vutha}}, \bibinfo {author}
  {\bibfnamefont {A.~D.}\ \bibnamefont {West}}, \ and\ \bibinfo {author}
  {\bibnamefont {{The ACME Collaboration}}},\ }\href {\doibase
  10.1126/science.1248213} {\bibfield  {journal} {\bibinfo  {journal}
  {Science}\ }\textbf {\bibinfo {volume} {343}},\ \bibinfo {pages} {269}
  (\bibinfo {year} {2014})}\BibitemShut {NoStop}%
\bibitem [{\citenamefont {Budker}(2014)}]{Budker2014}%
  \BibitemOpen
  \bibfield  {author} {\bibinfo {author} {\bibfnamefont {D.}~\bibnamefont
  {Budker}},\ }\href@noop {} {\enquote {\bibinfo {title} {{GNOME DAMOP
  2014}},}\ } (\bibinfo {year} {2014})\BibitemShut {NoStop}%
\bibitem [{\citenamefont {Roberts}\ \emph
  {et~al.}(2014{\natexlab{a}})\citenamefont {Roberts}, \citenamefont {Stadnik},
  \citenamefont {Dzuba}, \citenamefont {Flambaum}, \citenamefont {Leefer},\
  and\ \citenamefont {Budker}}]{RobertsCosPRD2014}%
  \BibitemOpen
  \bibfield  {author} {\bibinfo {author} {\bibfnamefont {B.~M.}\ \bibnamefont
  {Roberts}}, \bibinfo {author} {\bibfnamefont {Y.~V.}\ \bibnamefont
  {Stadnik}}, \bibinfo {author} {\bibfnamefont {V.~A.}\ \bibnamefont {Dzuba}},
  \bibinfo {author} {\bibfnamefont {V.~V.}\ \bibnamefont {Flambaum}}, \bibinfo
  {author} {\bibfnamefont {N.}~\bibnamefont {Leefer}}, \ and\ \bibinfo {author}
  {\bibfnamefont {D.}~\bibnamefont {Budker}},\ }\href {\doibase
  10.1103/PhysRevD.90.096005} {\bibfield  {journal} {\bibinfo  {journal}
  {Physical Review D}\ }\textbf {\bibinfo {volume} {90}},\ \bibinfo {pages}
  {096005} (\bibinfo {year} {2014}{\natexlab{a}})}\BibitemShut {NoStop}%
\bibitem [{\citenamefont {Roberts}\ \emph
  {et~al.}(2014{\natexlab{b}})\citenamefont {Roberts}, \citenamefont {Stadnik},
  \citenamefont {Dzuba}, \citenamefont {Flambaum}, \citenamefont {Leefer},\
  and\ \citenamefont {Budker}}]{RobertsCosmic2014}%
  \BibitemOpen
  \bibfield  {author} {\bibinfo {author} {\bibfnamefont {B.~M.}\ \bibnamefont
  {Roberts}}, \bibinfo {author} {\bibfnamefont {Y.~V.}\ \bibnamefont
  {Stadnik}}, \bibinfo {author} {\bibfnamefont {V.~A.}\ \bibnamefont {Dzuba}},
  \bibinfo {author} {\bibfnamefont {V.~V.}\ \bibnamefont {Flambaum}}, \bibinfo
  {author} {\bibfnamefont {N.}~\bibnamefont {Leefer}}, \ and\ \bibinfo {author}
  {\bibfnamefont {D.}~\bibnamefont {Budker}},\ }\href {\doibase
  10.1103/PhysRevLett.113.081601} {\bibfield  {journal} {\bibinfo  {journal}
  {Phys. Rev. Lett.}\ }\textbf {\bibinfo {volume} {113}},\ \bibinfo {pages}
  {81601} (\bibinfo {year} {2014}{\natexlab{b}})}\BibitemShut {NoStop}%
\bibitem [{\citenamefont {Abel}\ \emph {et~al.}(2017)\citenamefont {Abel},
  \citenamefont {Ayres}, \citenamefont {Ban}, \citenamefont {Bison},
  \citenamefont {Bodek}, \citenamefont {Bondar}, \citenamefont {Daum},
  \citenamefont {Fairbairn}, \citenamefont {Flambaum}, \citenamefont
  {Geltenbort}, \citenamefont {Green}, \citenamefont {Griffith}, \citenamefont
  {van~der Grinten}, \citenamefont {Gruji{\'{c}}}, \citenamefont {Harris},
  \citenamefont {Hild}, \citenamefont {Iaydjiev}, \citenamefont {Ivanov},
  \citenamefont {Kasprzak}, \citenamefont {Kermaidic}, \citenamefont {Kirch},
  \citenamefont {Koch}, \citenamefont {Komposch}, \citenamefont {Koss},
  \citenamefont {Kozela}, \citenamefont {Krempel}, \citenamefont {Lauss},
  \citenamefont {Lefort}, \citenamefont {Lemi{\`{e}}re}, \citenamefont {Marsh},
  \citenamefont {Mohanmurthy}, \citenamefont {Mtchedlishvili}, \citenamefont
  {Musgrave}, \citenamefont {Piegsa}, \citenamefont {Pignol}, \citenamefont
  {Rawlik}, \citenamefont {Rebreyend}, \citenamefont {Ries}, \citenamefont
  {Roccia}, \citenamefont {Rozp{\c{e}}dzik}, \citenamefont
  {Schmidt-Wellenburg}, \citenamefont {Severijns}, \citenamefont {Shiers},
  \citenamefont {Stadnik}, \citenamefont {Weis}, \citenamefont {Wursten},
  \citenamefont {Zejma},\ and\ \citenamefont {Zsigmond}}]{Abel2017}%
  \BibitemOpen
  \bibfield  {author} {\bibinfo {author} {\bibfnamefont {C.}~\bibnamefont
  {Abel}}, \bibinfo {author} {\bibfnamefont {N.~J.}\ \bibnamefont {Ayres}},
  \bibinfo {author} {\bibfnamefont {G.}~\bibnamefont {Ban}}, \bibinfo {author}
  {\bibfnamefont {G.}~\bibnamefont {Bison}}, \bibinfo {author} {\bibfnamefont
  {K.}~\bibnamefont {Bodek}}, \bibinfo {author} {\bibfnamefont
  {V.}~\bibnamefont {Bondar}}, \bibinfo {author} {\bibfnamefont
  {M.}~\bibnamefont {Daum}}, \bibinfo {author} {\bibfnamefont {M.}~\bibnamefont
  {Fairbairn}}, \bibinfo {author} {\bibfnamefont {V.~V.}\ \bibnamefont
  {Flambaum}}, \bibinfo {author} {\bibfnamefont {P.}~\bibnamefont
  {Geltenbort}}, \bibinfo {author} {\bibfnamefont {K.}~\bibnamefont {Green}},
  \bibinfo {author} {\bibfnamefont {W.~C.}\ \bibnamefont {Griffith}}, \bibinfo
  {author} {\bibfnamefont {M.}~\bibnamefont {van~der Grinten}}, \bibinfo
  {author} {\bibfnamefont {Z.~D.}\ \bibnamefont {Gruji{\'{c}}}}, \bibinfo
  {author} {\bibfnamefont {P.~G.}\ \bibnamefont {Harris}}, \bibinfo {author}
  {\bibfnamefont {N.}~\bibnamefont {Hild}}, \bibinfo {author} {\bibfnamefont
  {P.}~\bibnamefont {Iaydjiev}}, \bibinfo {author} {\bibfnamefont {S.~N.}\
  \bibnamefont {Ivanov}}, \bibinfo {author} {\bibfnamefont {M.}~\bibnamefont
  {Kasprzak}}, \bibinfo {author} {\bibfnamefont {Y.}~\bibnamefont {Kermaidic}},
  \bibinfo {author} {\bibfnamefont {K.}~\bibnamefont {Kirch}}, \bibinfo
  {author} {\bibfnamefont {H.-C.}\ \bibnamefont {Koch}}, \bibinfo {author}
  {\bibfnamefont {S.}~\bibnamefont {Komposch}}, \bibinfo {author}
  {\bibfnamefont {P.~A.}\ \bibnamefont {Koss}}, \bibinfo {author}
  {\bibfnamefont {A.}~\bibnamefont {Kozela}}, \bibinfo {author} {\bibfnamefont
  {J.}~\bibnamefont {Krempel}}, \bibinfo {author} {\bibfnamefont
  {B.}~\bibnamefont {Lauss}}, \bibinfo {author} {\bibfnamefont
  {T.}~\bibnamefont {Lefort}}, \bibinfo {author} {\bibfnamefont
  {Y.}~\bibnamefont {Lemi{\`{e}}re}}, \bibinfo {author} {\bibfnamefont
  {D.~J.~E.}\ \bibnamefont {Marsh}}, \bibinfo {author} {\bibfnamefont
  {P.}~\bibnamefont {Mohanmurthy}}, \bibinfo {author} {\bibfnamefont
  {A.}~\bibnamefont {Mtchedlishvili}}, \bibinfo {author} {\bibfnamefont
  {M.}~\bibnamefont {Musgrave}}, \bibinfo {author} {\bibfnamefont {F.~M.}\
  \bibnamefont {Piegsa}}, \bibinfo {author} {\bibfnamefont {G.}~\bibnamefont
  {Pignol}}, \bibinfo {author} {\bibfnamefont {M.}~\bibnamefont {Rawlik}},
  \bibinfo {author} {\bibfnamefont {D.}~\bibnamefont {Rebreyend}}, \bibinfo
  {author} {\bibfnamefont {D.}~\bibnamefont {Ries}}, \bibinfo {author}
  {\bibfnamefont {S.}~\bibnamefont {Roccia}}, \bibinfo {author} {\bibfnamefont
  {D.}~\bibnamefont {Rozp{\c{e}}dzik}}, \bibinfo {author} {\bibfnamefont
  {P.}~\bibnamefont {Schmidt-Wellenburg}}, \bibinfo {author} {\bibfnamefont
  {N.}~\bibnamefont {Severijns}}, \bibinfo {author} {\bibfnamefont
  {D.}~\bibnamefont {Shiers}}, \bibinfo {author} {\bibfnamefont {Y.~V.}\
  \bibnamefont {Stadnik}}, \bibinfo {author} {\bibfnamefont {A.}~\bibnamefont
  {Weis}}, \bibinfo {author} {\bibfnamefont {E.}~\bibnamefont {Wursten}},
  \bibinfo {author} {\bibfnamefont {J.}~\bibnamefont {Zejma}}, \ and\ \bibinfo
  {author} {\bibfnamefont {G.}~\bibnamefont {Zsigmond}},\ }\href {\doibase
  10.1103/PhysRevX.7.041034} {\bibfield  {journal} {\bibinfo  {journal} {Phys.
  Rev. X}\ }\textbf {\bibinfo {volume} {7}},\ \bibinfo {pages} {041034}
  (\bibinfo {year} {2017})}\BibitemShut {NoStop}%
\bibitem [{\citenamefont {Bovy}\ and\ \citenamefont
  {Tremaine}(2012)}]{Bovy:2012tw}%
  \BibitemOpen
  \bibfield  {author} {\bibinfo {author} {\bibfnamefont {J.}~\bibnamefont
  {Bovy}}\ and\ \bibinfo {author} {\bibfnamefont {S.}~\bibnamefont
  {Tremaine}},\ }\href {\doibase 10.1088/0004-637X/756/1/89} {\bibfield
  {journal} {\bibinfo  {journal} {Astrophys. J.}\ }\textbf {\bibinfo {volume}
  {756}},\ \bibinfo {pages} {89} (\bibinfo {year} {2012})},\ \Eprint
  {http://arxiv.org/abs/1205.4033} {arXiv:1205.4033} \BibitemShut {NoStop}%
\bibitem [{\citenamefont {Hern{\'{a}}ndez}\ \emph {et~al.}(2004)\citenamefont
  {Hern{\'{a}}ndez}, \citenamefont {Matos}, \citenamefont {Sussman},\ and\
  \citenamefont {Verbin}}]{Hernandez2004}%
  \BibitemOpen
  \bibfield  {author} {\bibinfo {author} {\bibfnamefont {X.}~\bibnamefont
  {Hern{\'{a}}ndez}}, \bibinfo {author} {\bibfnamefont {T.}~\bibnamefont
  {Matos}}, \bibinfo {author} {\bibfnamefont {R.~A.}\ \bibnamefont {Sussman}},
  \ and\ \bibinfo {author} {\bibfnamefont {Y.}~\bibnamefont {Verbin}},\ }\href
  {\doibase 10.1103/PhysRevD.70.043537} {\bibfield  {journal} {\bibinfo
  {journal} {Phys. Rev. D}\ }\textbf {\bibinfo {volume} {70}},\ \bibinfo
  {pages} {043537} (\bibinfo {year} {2004})}\BibitemShut {NoStop}%
\bibitem [{\citenamefont {Gonz{\'{a}}lez-Morales}\ \emph
  {et~al.}(2013)\citenamefont {Gonz{\'{a}}lez-Morales}, \citenamefont
  {Valenzuela},\ and\ \citenamefont {Aguilar}}]{Gonzalez-Morales2013}%
  \BibitemOpen
  \bibfield  {author} {\bibinfo {author} {\bibfnamefont {A.~X.}\ \bibnamefont
  {Gonz{\'{a}}lez-Morales}}, \bibinfo {author} {\bibfnamefont {O.}~\bibnamefont
  {Valenzuela}}, \ and\ \bibinfo {author} {\bibfnamefont {L.~A.}\ \bibnamefont
  {Aguilar}},\ }\href {\doibase 10.1088/1475-7516/2013/03/001} {\bibfield
  {journal} {\bibinfo  {journal} {J. Cosmol. Astropart. Phys.}\ }\textbf
  {\bibinfo {volume} {2013}},\ \bibinfo {pages} {001} (\bibinfo {year}
  {2013})}\BibitemShut {NoStop}%
\bibitem [{\citenamefont {Angstmann}\ \emph {et~al.}(2004)\citenamefont
  {Angstmann}, \citenamefont {Dzuba},\ and\ \citenamefont
  {Flambaum}}]{Angstmann2004}%
  \BibitemOpen
  \bibfield  {author} {\bibinfo {author} {\bibfnamefont {E.~J.}\ \bibnamefont
  {Angstmann}}, \bibinfo {author} {\bibfnamefont {V.~A.}\ \bibnamefont
  {Dzuba}}, \ and\ \bibinfo {author} {\bibfnamefont {V.~V.}\ \bibnamefont
  {Flambaum}},\ }\href {\doibase 10.1103/PhysRevA.70.014102} {\bibfield
  {journal} {\bibinfo  {journal} {Phys. Rev. A}\ }\textbf {\bibinfo {volume}
  {70}},\ \bibinfo {pages} {014102} (\bibinfo {year} {2004})}\BibitemShut
  {NoStop}%
\bibitem [{\citenamefont {Dinh}\ \emph {et~al.}(2009)\citenamefont {Dinh},
  \citenamefont {Dunning}, \citenamefont {Dzuba},\ and\ \citenamefont
  {Flambaum}}]{Dinh2009}%
  \BibitemOpen
  \bibfield  {author} {\bibinfo {author} {\bibfnamefont {T.~H.}\ \bibnamefont
  {Dinh}}, \bibinfo {author} {\bibfnamefont {A.}~\bibnamefont {Dunning}},
  \bibinfo {author} {\bibfnamefont {V.~A.}\ \bibnamefont {Dzuba}}, \ and\
  \bibinfo {author} {\bibfnamefont {V.~V.}\ \bibnamefont {Flambaum}},\ }\href
  {\doibase 10.1103/PhysRevA.79.054102} {\bibfield  {journal} {\bibinfo
  {journal} {Phys. Rev. A}\ }\textbf {\bibinfo {volume} {79}},\ \bibinfo
  {pages} {054102} (\bibinfo {year} {2009})}\BibitemShut {NoStop}%
\bibitem [{\citenamefont {{B.~M.~Roberts (2018)}}()}]{code}%
  \BibitemOpen
  \bibfield  {author} {\bibinfo {author} {\bibnamefont {{B.~M.~Roberts
  (2018)}}},\ }\href@noop {} {\bibinfo  {journal} {Code publicly available
  from:
  \href{https://github.com/benroberts999/DM-ClockAsymmetry}{github.com/benroberts999/DM-ClockAsymmetry}}\
  }\BibitemShut {NoStop}%
\bibitem [{\citenamefont {{Jet Propulsion Laboratory}}()}]{JPLigsac}%
  \BibitemOpen
\bibfield  {journal} {  }\bibfield  {author} {\bibinfo {author} {\bibnamefont
  {{Jet Propulsion Laboratory}}},\ }\href
  {ftp://sideshow.jpl.nasa.gov/pub/jpligsac/} {\bibinfo  {journal}
  {ftp://sideshow.jpl.nasa.gov/pub/jpligsac/}\ }\BibitemShut {NoStop}%
\bibitem [{\citenamefont {Jean}\ and\ \citenamefont
  {Dach}(2016)}]{MurphyJPL2015}%
  \BibitemOpen
\bibfield  {journal} {  }\bibfield  {author} {\bibinfo {author} {\bibfnamefont
  {Y.}~\bibnamefont {Jean}}\ and\ \bibinfo {author} {\bibfnamefont
  {R.}~\bibnamefont {Dach}},\ }\href@noop {} {\emph {\bibinfo {title} {IGS
  Central Bureau and University of Bern; Bern Open Publishing}}},\ \bibinfo
  {type} {Tech. Rep.}\ (\bibinfo {year} {2016})\BibitemShut {NoStop}%
\bibitem [{\citenamefont {Wcis{\l}o}\ \emph {et~al.}(2018)\citenamefont
  {Wcis{\l}o}, \citenamefont {Ablewski}, \citenamefont {Beloy}, \citenamefont
  {Bilicki}, \citenamefont {Bober}, \citenamefont {Brown}, \citenamefont
  {Fasano}, \citenamefont {Ciury{\l}o}, \citenamefont {Hachisu}, \citenamefont
  {Ido}, \citenamefont {Lodewyck}, \citenamefont {Ludlow}, \citenamefont
  {McGrew}, \citenamefont {Morzy{\'{n}}ski}, \citenamefont {Nicolodi},
  \citenamefont {Schioppo}, \citenamefont {Sekido}, \citenamefont {{Le
  Targat}}, \citenamefont {Wolf}, \citenamefont {Zhang}, \citenamefont
  {Zjawin},\ and\ \citenamefont {Zawada}}]{Wciso2018}%
  \BibitemOpen
  \bibfield  {author} {\bibinfo {author} {\bibfnamefont {P.}~\bibnamefont
  {Wcis{\l}o}}, \bibinfo {author} {\bibfnamefont {P.}~\bibnamefont {Ablewski}},
  \bibinfo {author} {\bibfnamefont {K.}~\bibnamefont {Beloy}}, \bibinfo
  {author} {\bibfnamefont {S.}~\bibnamefont {Bilicki}}, \bibinfo {author}
  {\bibfnamefont {M.}~\bibnamefont {Bober}}, \bibinfo {author} {\bibfnamefont
  {R.}~\bibnamefont {Brown}}, \bibinfo {author} {\bibfnamefont
  {R.}~\bibnamefont {Fasano}}, \bibinfo {author} {\bibfnamefont
  {R.}~\bibnamefont {Ciury{\l}o}}, \bibinfo {author} {\bibfnamefont
  {H.}~\bibnamefont {Hachisu}}, \bibinfo {author} {\bibfnamefont
  {T.}~\bibnamefont {Ido}}, \bibinfo {author} {\bibfnamefont {J.}~\bibnamefont
  {Lodewyck}}, \bibinfo {author} {\bibfnamefont {A.~D.}\ \bibnamefont
  {Ludlow}}, \bibinfo {author} {\bibfnamefont {W.~F.}\ \bibnamefont {McGrew}},
  \bibinfo {author} {\bibfnamefont {P.}~\bibnamefont {Morzy{\'{n}}ski}},
  \bibinfo {author} {\bibfnamefont {D.}~\bibnamefont {Nicolodi}}, \bibinfo
  {author} {\bibfnamefont {M.}~\bibnamefont {Schioppo}}, \bibinfo {author}
  {\bibfnamefont {M.}~\bibnamefont {Sekido}}, \bibinfo {author} {\bibfnamefont
  {R.}~\bibnamefont {{Le Targat}}}, \bibinfo {author} {\bibfnamefont
  {P.}~\bibnamefont {Wolf}}, \bibinfo {author} {\bibfnamefont {X.}~\bibnamefont
  {Zhang}}, \bibinfo {author} {\bibfnamefont {B.}~\bibnamefont {Zjawin}}, \
  and\ \bibinfo {author} {\bibfnamefont {M.}~\bibnamefont {Zawada}},\ }\href
  {\doibase 10.1126/sciadv.aau4869} {\bibfield  {journal} {\bibinfo  {journal}
  {Sci. Adv.}\ }\textbf {\bibinfo {volume} {4}},\ \bibinfo {pages} {eaau4869}
  (\bibinfo {year} {2018})},\ \Eprint {http://arxiv.org/abs/1806.04762}
  {arXiv:1806.04762} \BibitemShut {NoStop}%
\bibitem [{\citenamefont {Centers}\ \emph {et~al.}(2019)\citenamefont
  {Centers}, \citenamefont {Blanchard}, \citenamefont {Conrad}, \citenamefont
  {Figueroa}, \citenamefont {Garcon}, \citenamefont {Gramolin}, \citenamefont
  {Kimball}, \citenamefont {Lawson}, \citenamefont {Pelssers}, \citenamefont
  {Smiga}, \citenamefont {Sushkov}, \citenamefont {Wickenbrock}, \citenamefont
  {Budker},\ and\ \citenamefont {Derevianko}}]{StochasticAmplitude2019}%
  \BibitemOpen
  \bibfield  {author} {\bibinfo {author} {\bibfnamefont {G.~P.}\ \bibnamefont
  {Centers}}, \bibinfo {author} {\bibfnamefont {J.~W.}\ \bibnamefont
  {Blanchard}}, \bibinfo {author} {\bibfnamefont {J.}~\bibnamefont {Conrad}},
  \bibinfo {author} {\bibfnamefont {N.~L.}\ \bibnamefont {Figueroa}}, \bibinfo
  {author} {\bibfnamefont {A.}~\bibnamefont {Garcon}}, \bibinfo {author}
  {\bibfnamefont {A.~V.}\ \bibnamefont {Gramolin}}, \bibinfo {author}
  {\bibfnamefont {D.~F.~J.}\ \bibnamefont {Kimball}}, \bibinfo {author}
  {\bibfnamefont {M.}~\bibnamefont {Lawson}}, \bibinfo {author} {\bibfnamefont
  {B.}~\bibnamefont {Pelssers}}, \bibinfo {author} {\bibfnamefont {J.~A.}\
  \bibnamefont {Smiga}}, \bibinfo {author} {\bibfnamefont {A.~O.}\ \bibnamefont
  {Sushkov}}, \bibinfo {author} {\bibfnamefont {A.}~\bibnamefont
  {Wickenbrock}}, \bibinfo {author} {\bibfnamefont {D.}~\bibnamefont {Budker}},
  \ and\ \bibinfo {author} {\bibfnamefont {A.}~\bibnamefont {Derevianko}},\
  }\href {http://arxiv.org/abs/1905.13650} {\  (\bibinfo {year} {2019})},\
  \Eprint {http://arxiv.org/abs/1905.13650} {arXiv:1905.13650} \BibitemShut
  {NoStop}%
\end{thebibliography}%

\end{document}